\input harvmac
\noblackbox
 
\font\ticp=cmcsc10
 
\def\Title#1#2{\rightline{#1}\ifx\answ\bigans\nopagenumbers\pageno0\vskip1in
\else\pageno1\vskip.8in\fi \centerline{\titlefont #2}\vskip .5in}

\font\ticp=cmcsc10
\font\ttsmall=cmtt10 at 8pt

\input epsf
\ifx\epsfbox\UnDeFiNeD\message{(NO epsf.tex, FIGURES WILL BE
IGNORED)}
\def\figin#1{\vskip2in}
\else\message{(FIGURES WILL BE INCLUDED)}\def\figin#1{#1}\fi
\def\ifig#1#2#3{\xdef#1{fig.~\the\figno}
\goodbreak\topinsert\figin{\centerline{#3}}%
\smallskip\centerline{\vbox{\baselineskip12pt
\advance\hsize by -1truein\noindent{\bf Fig.~\the\figno:} #2}}
\bigskip\endinsert\global\advance\figno by1}
%
%
\def\[{\left [}
\def\]{\right ]}
\def\({\left (}
\def\){\right )}
\def\p{\partial}

\def\S{{\bf S}}
\def\l{\ell}
\def\g{\gamma}
\def\CC{\{ \cal{C} \}}
\def\C{{\cal{C}}}
\def\d{\delta}
\def\e{\varepsilon}

\def\Om{\Omega}
\def\a{\alpha}
\def\b{\beta}

\def\CN{{\cal N}}

\def\CI{I}
\def\scri{{\cal I}}

\def\Ci{i}
\def\ud{\dot{u}}
\def\vd{\dot{v}}
\def\xd{\dot{x}}
\def\xid{\dot{x^i}}
\def\fd{\dot{f}}
\def\rd{\dot{r}}

\def\rd{\dot{r}}
\def\vx{{\vec x}}
\def\dua{\( {\p \over \p u} \)^a}
\def\dva{\( {\p \over \p v} \)^a}
\def\dxa{\( {\p \over \p x^i} \)^a}
\def\vdd{\ddot{v}}
\def\xdd{\ddot{x}}
\def\ptd{(u\!=\!-\delta \, ,\, v\!=\!0 \, ,\, x^i\!=\!0)}
\def\itm{\noindent $\bullet$ \ }

\def\eg{{\it e.g.}}
\def\ie{{\it i.e.}}
\def\cf{{\it cf.}}
\def\vz{{\vec z}}
\def\M{{\cal{M}}}
\def\P{{\cal{P}}}


\lref\maro{
D.~Marolf and S.~F.~Ross,
{\it Plane waves: To infinity and beyond!},
[arXiv:hep-th/0208197].}

\lref\geroch{R.~Geroch, E.~H.~Kronheimer, and R.Penrose,
{\it Ideal points in space-time},
Proc.\ Roy.\ Soc.\ Lond.\ A.\ {\bf 327}, 545 (1972).}

\lref\penrose{
R.~Penrose,
``Any spacetime has a planewave as a limit,'' in
{\it Differential geometry and relativity}, pp 271-275,
Reidel, Dordrecht, 1976.}

\lref\penr{
R.~Penrose,
{\it A Remarkable Property Of Plane Waves In General Relativity},
Rev.\ Mod.\ Phys.\  {\bf 37}, 215 (1965).}

\lref\bmn{
D.~Berenstein, J.~M.~Maldacena and H.~Nastase,
{\it Strings in flat space and pp waves from N = 4 super Yang Mills},
JHEP {\bf 0204}, 013 (2002)
[arXiv:hep-th/0202021].}

\lref\host{
G.~T.~Horowitz and A.~R.~Steif,
{\it Space-Time Singularities In String Theory},
Phys.\ Rev.\ Lett.\  {\bf 64}, 260 (1990).}

\lref\bfhp{
M.~Blau, J.~Figueroa-O'Farrill, C.~Hull and G.~Papadopoulos,
{\it A new maximally supersymmetric background of IIB superstring theory},
JHEP {\bf 0201}, 047 (2002)
[arXiv:hep-th/0110242].}

\lref\bena{
D.~Berenstein and H.~Nastase,
{\it On light-cone string field theory from super Yang-Mills and holography},
[arXiv:hep-th/0205048].}

\lref\tseytlin{
A.~A.~Tseytlin,
{\it On limits of superstring in AdS(5) x S**5},
[arXiv:hep-th/0201112].}

\lref\juan{
J. Maldacena, 
{\it The Large N Limit of Superconformal Field Theories and Supergravity},
Adv. Theor. Math. Phys. {\bf 2} (1998) 231, [arXiv:hep-th/9711200].}

\lref\magoo{
O. Aharony, S.S. Gubser, J. Maldacena, H. Ooguri, Y. Oz,
{\it Large N Field Theories, String Theory and Gravity},
Phys. Rept. {\bf 323} (2000) 183, [arXiv:hep-th/9905111].}

\lref\witten{ 
E. Witten, 
{\it Anti De Sitter Space And Holography},
Adv. Theor. Math. Phys. {\bf 2} (1998) 253, [arXiv:hep-th/9802150].}

\lref\gkp{
S. Gubser, I. Klebanov, and A. Polyakov, 
{\it Gauge Theory Correlators from Non-Critical String Theory},
Phys. Lett. {\bf B428} (1998) 105,
[arXiv:hep-th/9802109].}

\lref\bfp{
M.~Blau, J.~Figueroa-O'Farrill and G.~Papadopoulos,
{\it Penrose limits, supergravity and brane dynamics},
[arXiv:hep-th/0201081].}

\lref\amkl{
D.~Amati and C.~Klimcik,
{\it Nonperturbative Computation Of The Weyl Anomaly For A Class Of 
Nontrivial Backgrounds},
Phys.\ Lett.\ B {\bf 219}, 443 (1989).}

\lref\nobh{
V.~E.~Hubeny and M.~Rangamani,
{\it No black holes in pp-waves},
[arXiv:hep-th/0210234]}

\lref\metsaev{
R.~R.~Metsaev,
{\it Type IIB Green-Schwarz superstring in plane wave 
Ramond-Ramond  background},
Nucl.\ Phys.\ B {\bf 625}, 70 (2002)
[arXiv:hep-th/0112044].}

\lref\trusso{
J.~G.~Russo and A.~A.~Tseytlin,
{\it A class of exact pp-wave string models with interacting 
light-cone  gauge actions},
JHEP {\bf 0209}, 035 (2002)
[arXiv:hep-th/0208114].}

\lref\malmaoz{
J.~Maldacena and L.~Maoz,
{\it Strings on pp-waves and massive two dimensional field theories},
[arXiv:hep-th/0207284].}

\lref\mrtwo{
D.~Marolf and S.~Ross,
{\it To appear}.}

\lref\don{
D.~Marolf,
{\it Private communication}.}

\lref\prec{
V.~E.~Hubeny,
{\it Precursors see inside black holes},
[arXiv:hep-th/0208047].}

\lref\nonlocal{
V.~E.~Hubeny, M.~Rangamani and E.~Verlinde,
{\it Penrose limits and non-local theories}, 
JHEP {\bf 0210}, 020 (2002)
[arXiv:hep-th/0205258].}

\lref\googuri{
J.~Gomis and H.~Ooguri,
{\it Penrose limit of N = 1 gauge theories},
Nucl.\ Phys.\ B {\bf 635}, 106 (2002)
[arXiv:hep-th/0202157].}

\lref\candela{
A.~M.~Candela, J.~L.~Flores and M.~Sanchez,
{\it On general plane fronted waves. Geodesics,}
[arXiv:gr-qc/0211017.]}

\lref\bicak{
J.~Bi\v{c}\'{a}k,
{\it The role of exact solutions of Einstein's equations in the  developments
 of general relativity and astrophysics selected themes,}
Lect.\ Notes Phys.\  {\bf 540}, 1 (2000)
[arXiv:gr-qc/0004016].}

\lref\ek{J.~Ehlers and K.~Kundt,
{\it Exact solutions of the gravitational field equations, in Gravitation: 
an introduction to current research},
 ed. L.~Witten, J.~Wiley\&Sons, New York (1962)}

\lref\gili{S.~B.~Giddings and M.~Lippert,
{\it Precursors, black holes, and a locality bound},
Phys.\ Rev.\ D 65, 024006 (2002),
[arXiv:hep-th/0103231].}

\lref\kali{
D.~Kabat and G.~Lifschytz,
{\it Gauge theory origins of supergravity causal structure},
JHEP 9905, 005 (1999), [arXiv:hep-th/9902073].}

\lref\maldacena{J.~M.~Maldacena,
{\it Eternal black holes in Anti-de-Sitter},
[arXiv:hep-th/0106112].}

\lref\jacobson{
T.~Jacobson,
{\it On the nature of black hole entropy,}
[arXiv:gr-qc/9908031].}

\lref\lmr{
J. Louko, D. Marolf, and S. F. Ross,
{\it On geodesic propagators and black hole holography},
Phys.Rev. D62 (2000) 044041, [arXiv:hep-th/0002111].}

\lref\gradr{
I.~S.~Gradshteyn and I.~M.~Ryzhik,
{\it Table of Integrals, Series, and Products},
Fifth Ed., ed. A.~Jefferey, Academic Press, London (1994).
}

\lref\napwit{
C.~R.~Nappi and E.~Witten,
{\it A WZW model based on a nonsemisimple group},
Phys.\ Rev.\ Lett.\  {\bf 71}, 3751 (1993)
[arXiv:hep-th/9310112].}

\lref\akisunny{
A.~Hashimoto and N.~Itzhaki,
{\it Non-commutative Yang-Mills and the AdS/CFT correspondence},
Phys.\ Lett.\ B {\bf 465}, 142 (1999)
[arXiv:hep-th/9907166].}

\lref\mrusso{
J.~M.~Maldacena and J.~G.~Russo,
{\it Large N limit of non-commutative gauge theories},
JHEP {\bf 9909}, 025 (1999)
[arXiv:hep-th/9908134].}

\lref\kirit{
E.~Kiritsis and B.~Pioline,
{\it Strings in homogeneous gravitational waves and null holography},
JHEP {\bf 0208}, 048 (2002)
[arXiv:hep-th/0204004].}

\lref\martineca{
E.~J.~Martinec,
{\it Strings And Causality},
[arXiv:hep-th/9311129].}

\lref\martinecb{
E.~J.~Martinec,
{\it The Light cone in string theory},
Class.\ Quant.\ Grav.\  {\bf 10}, L187 (1993)
[arXiv:hep-th/9304037].}

\lref\mpz{
D.~Marolf and L.~A.~Zayas,
{\it On the singularity structure and stability of plane waves},
[arXiv:hep-th/0210309].}

\lref\shamit{
S.~Kachru and L.~McAllister,
{\it Bouncing brane cosmologies from warped string compactifications},
[arXiv:hep-th/0205209].}

\lref\craps{
B.~Craps, D.~Kutasov and G.~Rajesh,
{\it String propagation in the presence of cosmological singularities},
JHEP {\bf 0206}, 053 (2002)
[arXiv:hep-th/0205101].}

\lref\cornalbac{
L.~Cornalba, M.~S.~Costa and C.~Kounnas,
{\it A resolution of the cosmological singularity with orientifolds},
Nucl.\ Phys.\ B {\bf 637}, 378 (2002)
[arXiv:hep-th/0204261].}

\lref\vijay{
V.~Balasubramanian, S.~F.~Hassan, E.~Keski-Vakkuri and A.~Naqvi,
{\it A space-time orbifold: A toy model for a cosmological singularity},
[arXiv:hep-th/0202187].}

\lref\horpol{
G.~T.~Horowitz and J.~Polchinski,
{\it Instability of spacelike and null orbifold singularities},
[arXiv:hep-th/0206228].}

\lref\lmsb{
H.~Liu, G.~Moore and N.~Seiberg,
{\it Strings in time-dependent orbifolds},
JHEP {\bf 0210}, 031 (2002)
[arXiv:hep-th/0206182].}

\lref\lmsa{
H.~Liu, G.~Moore and N.~Seiberg,
{\it Strings in a time-dependent orbifold},
JHEP {\bf 0206}, 045 (2002)
[arXiv:hep-th/0204168].}

\lref\albion{
A.~Lawrence,
{\it On the instability of 3D null singularities},
JHEP {\bf 0211}, 019 (2002)
[arXiv:hep-th/0205288].}

\lref\michal{
M.~Fabinger and J.~McGreevy,
{\it On smooth time-dependent orbifolds and null singularities},
arXiv:hep-th/0206196.}
 
\lref\andy{
A.~Strominger,
{\it The dS/CFT correspondence},
JHEP {\bf 0110}, 034 (2001)
[arXiv:hep-th/0106113].}

\lref\bousso{
R.~Bousso,
{\it The holographic principle},
Rev.\ Mod.\ Phys.\  {\bf 74}, 825 (2002)
[arXiv:hep-th/0203101].}

\lref\bgs{
D.~Brecher, J.~P.~Gregory and P.~M.~Saffin,
{\it String theory and the classical stability of plane waves},
[arXiv:hep-th/0210308].
}

%
\baselineskip 16pt
\Title{\vbox{\baselineskip12pt
\line{\hfil SU-ITP-02/43}
\line{\hfil UCB-PTH-02/51}
\line{\hfil LBNL-51744 }
\line{\hfil \tt hep-th/0211195} }}
{\vbox{
{\centerline{Causal structures of pp-waves}
}}}
\centerline{\ticp Veronika E. Hubeny$^a$
 and Mukund Rangamani$^{b,c}$ \footnote{}{\ttsmall
 veronika@itp.stanford.edu, mukund@socrates.berkeley.edu}}
\bigskip
\centerline {\it $^a$
Department of Physics, Stanford University, Stanford, CA 94305, USA} 
\centerline{\it $^b$ Department of Physics, University of California,
Berkeley, CA 94720, USA} 
\centerline{\it $^c$ Theoretical Physics Group, LBNL, Berkeley, CA 94720, USA}

\bigskip
\centerline{\bf Abstract}
\bigskip

We discuss the causal structure of pp-wave spacetimes using the 
ideal point construction outlined by Geroch, Kronheimer, and Penrose. 
This generalizes the recent work of Marolf and Ross, who considered 
similar issues for plane wave spacetimes. 
We address the question regarding the dimension of the causal 
boundary for certain specific pp-wave backgrounds. In particular, 
we demonstrate that the pp-wave spacetime 
which gives rise to the $ \CN = 2$  sine-Gordon string 
world-sheet theory is geodesically complete and has a 
one-dimensional causal boundary.

\Date{November, 2002}
%
\newsec{Introduction}

{\it pp-waves} (or ``plane-fronted waves with parallel rays'')
are all spacetimes with covariantly constant null Killing field.
In general relativity, they form simple solutions to Einstein's equations 
with many curious properties. The presence of the covariantly constant 
null Killing field implies that these spacetimes have vanishing scalar 
curvature invariants, much the same as flat space. {\it Plane waves} are
a subset of these which have in addition an extra ``planar'' 
symmetry along the wavefronts. They can be thought of as arising 
from the so-called Penrose limit
\penrose\ of any spacetime, which essentially consists of zooming in
onto any null geodesic in that spacetime.  
Nevertheless, these are 
distinct from flat spacetime and their structure is much richer.
Interestingly, as shown by Penrose in \penr, 
plane wave spacetimes are not globally hyperbolic, 
so that there exists no Cauchy hypersurface from which a causal evolution 
would cover the entire spacetime.
This automatically implies that even the causal structure 
of pp-waves is different from that of flat spacetime.

pp-wave spacetimes are especially important within
the context of string theory.
This is because they yield exact classical backgrounds for 
string theory, since all curvature invariants, 
and therefore all $\alpha'$ corrections, vanish \refs{\amkl,\host}. 
Hence the pp-wave spacetimes correspond to exact conformal field theories.
Because of this fact, they provide much-needed examples of classical
solutions in string theory, which can in turn be used as toy models 
for studying its structure and properties. 
Plane waves happen to be even simpler, for the action in light-cone gauge
is quadratic.

While this fact has been appreciated for some time \refs{\amkl, \host},
only recently have plane waves received significant attention, 
mainly initiated by
the work of Berenstein, Maldacena, and Nastase (BMN) \bmn, based on
the AdS/CFT correspondence \refs{\juan,\witten,\gkp,\magoo}.
These authors
proposed a very interesting solvable model of string theory in 
Ramond-Ramond backgrounds  by taking the Penrose limit of 
$AdS_5 \times \S^5$ spacetime \refs{\bmn, \tseytlin}, the 
holographic dual of $d=4$, $\CN =4$ Super-Yang-Mills theory. 
This maximally supersymmetric plane wave solution of Type IIB 
supergravity \bfhp\ (henceforth BMN plane wave) happens to be the simplest 
example of a sigma model with 
Ramond-Ramond background that is solvable \metsaev.
Further developments in this area include other interesting solvable
or integrable world-sheet theories \refs{\malmaoz, \trusso}.

A very intriguing aspect of the BMN plane wave 
is that its conformal boundary happens to be a one-dimensional 
null line. This was first demonstrated in \bena\ using 
the standard technique of conformally mapping the plane wave spacetime 
into the Einstein Static Universe to construct the Penrose 
diagram. Later Marolf and Ross \maro\ showed the same using a 
more sophisticated technique of adding `points at infinity', a construction
dating back to the work of Geroch, Kronheimer and Penrose \geroch. 
The latter technique has the added advantage of being applicable 
for spacetimes that are not conformally flat as opposed to the plane wave 
solution arising from the Penrose limit of $AdS_5 \times \S^5$. 

In a preceding paper \nobh, we had asked a general question: 
{\it Do pp-waves admit event horizons?} The primary motivation for the 
same was to check whether there were black hole like spacetimes admitting 
a covariantly constant null Killing field. If the answer to the 
question were in the affirmative, we would have black hole solutions 
that remain exact conformal field theories to all orders in the 
perturbative $\a'$ expansion. In addition, one might hope to be 
able to delve deeper into the mysteries of 
black holes using perturbative string technology, provided that these 
solutions proved amenable to light-cone quantization.

In \nobh\  we have argued that plane waves cannot
admit event horizons, because every point of the spacetime can
communicate ``out to infinity''.  
Since black holes are defined as the regions bounded by event horizons,
this automatically shows that there can't be black hole pp-waves.

While the fact that the spacetime admits no event horizons 
gives us an important information about its causal structure, 
it of course does not determine this causal structure in its entirety.
One may well ask, why should we want to know this causal structure 
of a spacetime? 
The motivation for such analysis, apart from its obvious interest
to general relativity, is that the causal structure of a spacetime gives 
us some important information about the spacetime.
In particular, in the spirit of AdS/CFT correspondence, 
the structure of the boundary may hint at 
possible background on which a potential dual theory would live.
While it is certainly not guaranteed that there will be a dual
theory ``living on the  boundary'' of a spacetime,
knowledge of the causal structure might prove of some use in 
determining the same. Also knowledge of the causal structure 
enables one to make stronger arguments than those presented in 
\nobh\ regarding the question of the presence/absence of event horizons. 
By outlining the general properties of the causal structure that 
pp-wave spacetimes satisfy, we put our analysis at a level of greater 
robustness. 

An important issue that  the knowledge of the causal structure allows us to 
to discuss is the dimensionality of the causal boundary. Most 
typical $d$-dimensional spacetimes have a $d-1$-dimensional  boundary.
Such is true for all the standard examples, such as Minkowski,  
Anti-deSitter and deSitter spacetimes, and for more generic solutions 
which asymptote to the same, such as black holes in these spacetimes. 
The startling fact which is revealed by the analysis of \bena, \maro\ is 
that for the maximally supersymmetric plane wave solution of Type IIB 
supergravity the  boundary is one-dimensional! In fact, the 
same is true for certain other classes of plane waves such as those 
arising from the Penrose limits of $AdS_7 \times \S^4$,
 $AdS_4 \times \S^7$ and the near horizon geometry of D4 branes \maro, 
among others. There however, are examples of plane wave spacetimes where 
the  boundary isn't one-dimensional. We wish to ask whether 
the pp-wave spacetimes share similar properties and will argue for 
some classes of pp-waves (including some that lead to integrable 
world-sheet theories) that the causal boundary\foot{
There is a slight subtlety relating to the distinction between causal 
boundary and conformal boundary, having to do with the topology of the
completed manifold.  While we mention this at the end of Section 3, 
in the present work we bypass these topological subtleties by confining
our discussion to causal structures, as was done by \maro.
} is indeed one-dimensional.

The outline of this paper is as follows.
In the following short section, we review certain basic aspects of 
plane wave and pp-wave spacetimes, mainly with the view of setting 
up notation. In Section 3, we review the work of Geroch, Kronheimer 
and Penrose, providing the ingredients necessary for determining the 
causal structure of any spacetime. We then turn to the question of 
causal structure of general plane waves in Section 4, 
reviewing the arguments of 
Marolf and Ross, and comment on some generalizations, and 
present a few examples. 
In Section 5, we turn to discussing the causal structure of 
general pp-waves and construct the causal structure for certain interesting 
pp-wave solutions. 
We end in Section 6 with a brief summary and more general discussion
of singularities and causal structure.
In Appendix A we collect some useful facts about plane and pp-wave
spacetimes, and in Appendix B we present details relating to null 
geodesics in particular vacuum pp-waves.

\newsec{Notation and terminology}

To set the notation and re-emphasize terminology, we will write explicitly
three classes of spacetimes, in decreasing generality.
The {\it pp-wave} spacetimes, 
which are defined as all spacetimes admitting a covariantly constant null
Killing field,  can be written as 
\eqn\ppgen{
ds^2 = -2 \, du \, dv - F(u, x^i) \, du^2 + A_i(u,x^i) \, du \, dx^i + 
g_{ij}(x^i) \, dx^i \, dx^j .}
We shall in the following be working with spacetimes wherein $A(u,x^i) = 0$
in order to maintain the simplicity of the discussion.  
For the case of pure gravity, 
vacuum Einstein's equations dictate that $F(u,x^i)$ satisfy
the transverse Laplace equation for each $u$ and that the transverse space be 
Ricci flat. $F(u,x^i)$, however, can be an arbitrary function of $u$. 
Another simplification that we will make is to consider pp-wave spacetimes 
with flat transverse part, { \it i.e.}, we will consider spacetimes with the 
metric
\eqn\pp{
ds^2 = -2 \, du \, dv - F(u, x^i) \, du^2 +  dx^i \, dx^i .}

{\it Plane wave} spacetimes are those where the harmonic function in \pp\
 is in fact quadratic, $F(u, x^i) = f_{ij}(u) \, x^i x^j$,
 so that plane waves can be written as
\eqn\plane{
ds^2 = -2 \, du \, dv - f_{ij}(u) \, x^i x^j \, du^2 + dx^i \, dx^i}
Here, $f_{ij}(u)$ can be any function of $u$, subject to the constraint
that for each $u$,
$f_{ij}$ is symmetric and traceless (the latter being required by
vacuum Einstein's equations).  As suggested by the name, these metrics
have an extra ``plane'' symmetry, which contains the translations
along the wave-fronts in the transverse directions.  This can be seen
explicitly by casting \plane\ into the Rosen form,\foot{
Typically, this metric is not geodesically complete because of 
coordinate singularities, but the Brinkman form \plane\ does 
cover the full spacetime.
The coordinate transformation from one form into the other is given
e.g.\ in \bfp.
For metric of the Brinkman form
 $ds^2 = - 2 \, du \, dv - f(u) \, x^2 \, du^2 + dx^2$, the
coordinate transformation 
$\{ u= U, x=  h(U)\, X , v= V+ {1 \over 2} h(U) \, h'(U) \, X^2 \}$
where $h(U)$ satisfies $h''(U) + f(U) \, h(U) = 0$,
casts this metric into the Rosen form
$ds^2 = -2 \, dU \, dV + h(U)^2 \, dX^2$.}
\eqn\rosen{
ds^2 = -2 \, dU \, dV + C_{ij}(U) \, dX^i \, dX^j}
The {\it homogeneous plane waves} further specialize \plane\ 
by taking out $f$'s dependence on $u$,
\eqn\homplane{
ds^2 = -2 \, du \, dv - f_{ij} \, x^i x^j \, du^2 + dx^i \, dx^i}
The BMN plane wave metric \bmn, 
found earlier by \bfhp, belongs to this last class, for the special 
case $f_{ij} = \mu^2 \, \delta_{ij}$, and $u \equiv x^+, v \equiv x^-$ 
in their notation.  In fact, in the constant $f$ case, we can diagonalize
the metric completely, which leads to substantial simplification in 
the analysis, as used \eg\ by \maro.

All the aforementioned spacetimes have a covariantly constant 
null Killing vector, given by $p^a = \( {\p \over \p v }\)^a$.
The fact that this is a null Killing vector is obvious from the metric, 
while its being covariantly constant may be inferred from the vanishing 
of the Christoffel symbols $\Gamma^v_{\ \mu v}$. 

\newsec{Causal structure generalities}

The most conventional, and often the easiest, way to determine
the causal structure of a spacetime is to conformally map the 
spacetime into the Einstein Static Universe (ESU), and see where the
conformal factor diverges.   
This method was employed {\it e.g.}\ by \bena\ to determine the asymptotic
structure of the BMN plane wave.
This approach, however, only works for a limited class of spacetimes,
for which a conformal factor exists, {\it i.e.}, conformally flat ones
(defined by the vanishing of the Weyl tensor).
As pointed out by \maro, this condition is not satisfied by general
plane waves (nor by the more general pp-waves). 
In fact, the Weyl tensor for the metric
\pp\ in $d$ dimensional spacetime is given by 
\eqn\weylpp{
C_{uiuj} = {1 \over 2} \( \p_i \p_j F(u, x) 
            - {1 \over d-2} \, \delta_{ij} \,
              \sum_k \p_k^2 F(u, x) \) }
where $i,j = 1, \dots, d-2$,
which, in the plane wave spacetimes \plane\ reduces to
\eqn\weyl{
C_{uiuj} = f_{ij}(u) - {1 \over d-2} \, \delta_{ij} \,
\sum_k f_{kk}(u).}

Since the only nonzero component of the Ricci tensor is
\eqn\ricci{
R_{uu} = {1 \over 2 } \nabla^2_T \, F(u,x^i)}
and the Ricci scalar then automatically vanishes, 
the only vacuum pp-wave which is conformally flat is the 
trivial (flat spacetime) one, where 
$F(u,x^i) = a + b_i \, x^i$, which can be cast into the form
\pp\ with $F(u,x^i) \equiv 0$ by appropriate coordinate transformation.
On the other hand, certain nontrivial plane (or pp) waves can be 
conformally flat if we allow fluxes, such as in the BMN plane
wave, where $F(u,x^i) = \mu^2 \, \sum_i \, (x^i)^2 $.
For this class of spacetimes one may use the ESU conformal mapping 
procedure to determine the causal structure.  Unfortunately, 
this class of spacetimes is rather limited; for instance, the Penrose
limits of $AdS_4 \times \S^7$ and $AdS_7 \times \S^4$ analogous to the
BMN plane wave do {\it not} fall into this category. Similar 
considerations hold for the case of general pp-waves.

To bypass that obstacle, \maro\ used a more direct approach to find the
causal structure of these spacetimes, based on the method introduced
by Geroch, Kronheimer, and Penrose \geroch.
In the present section, we will present a self-contained review of
the method of ideal point construction.

\subsec{Review of ideal point construction}

To find the causal structure of a given spacetime, 
\geroch\ complete the spacetime by  ``ideal points'',
corresponding, roughly speaking, to the endpoints of inextendible
causal curves $\g$. 
This procedure of adding the boundary to our spacetime allows us
to apply the usual notions of causality to this boundary, thereby
extracting the causal structure.
We will first develop the necessary terminology and then outline
the prescription in more detail.

Let $\CI^-(p)$ denote the indecomposable past-set (IP) of a
given point $p$ in the spacetime. This is basically the collection of 
points from which there exists a future-directed causal curve\foot{
We define {\it causal} curves to be timelike or null, so that
unlike the usual convention, we take the set $\CI^-(p)$ to be closed.
This makes our notation and arguments cleaner; however, we could have
taken the stricter definition of IPs defined by timelike curves only, 
and used strict inequalities correspondingly.}
 to $p$,
 or in other words, the points lying in the past light-cone of $p$. 
We can consider instead of a point a future-directed causal curve 
 itself, and denote 
by $\CI^-\[\g \]$ the IP associated with the curve $\g$. Of course, 
this is simply the union of the IPs associated with all the points lying 
on the curve $\g$, {\it i.e.}, $\CI^-\[ \g \] = \bigcup_i \, \CI^-\(p_i\)$ 
for all $p_i \in \gamma$. Now an IP is a proper IP or PIP 
if there exists some point $p$ in the manifold such that $I^-[\gamma] 
= \CI^-\( p \)$; else the IP is a terminal IP or TIP. TIPs can be seen to 
correspond to future endless causal curves. 
We can extend the definitions to past-directed causal curves,
thereby creating indecomposable future-set (IF) and thence PIF and TIF 
respectively. Note that the curves can be endless either because they run off 
to infinity, or because they hit a physical singularity (since singularities 
are by definition not a part of our physical spacetime). Hence the set of 
ideal points will describe both the causal boundary of our spacetime, as 
well any singularities. 

To construct the causal structure of a spacetime, the prescription is 
as follows:

\noindent
1. First find the set of TIP(F)s, or ``terminal indecomposable past (future)
 sets'', $\CI^{\pm}\[ \g \]$, of all causal curves $\g$. 
Since curves corresponding to TIP(F)s  are future (past) endless, we can
add an ``ideal point'' to our physical spacetime.
This is a basically a point at infinity, such that 
the TIPs of the curves would be converted into PIPs were this 
ideal point part of the spacetime manifold.
The ideal points are thus added in by hand to the original manifold and 
one has two completions of the original manifold $M$,
$\hat{M}$ and $\check{M}$ corresponding to the original manifold with 
all the ideal points given by TIPs and TIFs respectively.
This, however, is not the entire story: whereas we have added points
on which causal curves end, we haven't yet specified the relations
between these added ideal points, which brings us to the second element
of the prescription.

\noindent 
2. Consider the manifold $M^{\sharp} = \hat{M} \cup \check{M}$. The 
global completion of $M$ to a manifold with a nice conformal boundary is 
defined to be $\bar{M} = M^\sharp / \Gamma$ where $\Gamma$  is the minimal 
set of identifications between the ideal points thus added,
 such that $\bar{M}$ is a smooth Hausdorff manifold. 

\ifig\figCaus{Example of identifications between TIPs and TIFs.
(a) The causal curves $g1$ and $g2$ have the same TIP;
(b) Similarly, the causal curves $g3$ and $g4$ have the same TIF;
(c) Finally, the TIF of $g5$ and TIP of $g6$ are identified.
In each case, there is correspondingly a single ideal point 
(labeled $P$, $Q$, and $R$, respectively) attached.}
{\epsfxsize=10cm \epsfysize=6cm \epsfbox{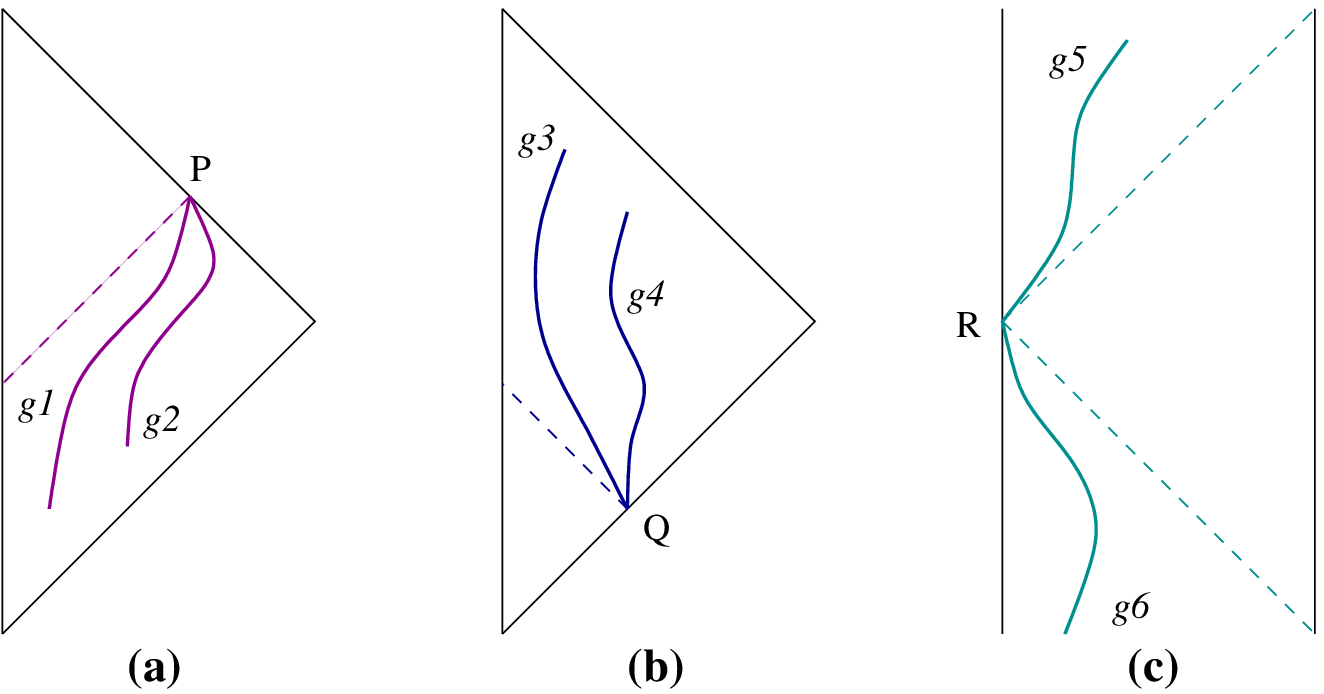}}

The second point, concerning identifications, deserves a bit more
explanation.
Since the ideal points were defined through the TIP(F)s, distinct 
ideal points should correspondingly have distinct TIP(F)s.
In other words, different causal curves may nevertheless have identical
past and/or future, so that their TIPs and/or TIFs are identical sets; 
 if such is the case, the corresponding ideal points are identified.  
This is illustrated by the examples in Fig.1:
In Fig.1(a), the causal curves $g1$ and $g2$ have the same past, lying
below the dashed line.  Correspondingly, we attach only a single ideal
point, $P$, to their TIP.  Similarly, in Fig.1(b),
there is only a single ideal point $Q$ corresponding to the TIF of the
causal curves $g3$ and $g4$. 
A more subtle (but very important) identification is illustrated 
in Fig.1(c), where the past endpoint of curve $g5$, $R$, is the 
future endpoint of a curve $g6$.
Thus, we see that in this case,
 some TIPs may need to be identified with some TIFs.
This necessarily occurs whenever the boundary is timelike, such as for
AdS; but we will see below that it can in fact occur for null boundaries
as well. 

Before proceeding, we should put in a side cautionary remark.
An important feature associated with the identifications of the 
ideal points has to do with the topology of the resulting completed
spacetime. 
One expects on physical grounds that conformally completed
spacetime manifold has a Hausdorff topology, enabling one to 
distinguish between distinct points. If one uses the GKP construction to 
causally complete the spacetime with ideal points, the completed
manifold is by definition a Hausdorff spacetime; so one
might expect that the conformal completion of the spacetime 
should also have the requisite Hausdorff topology. 
However, this may be in general a more subtle issue than 
the simple set of identifications we discuss above, \ie\ based on the causal 
properties associated with the ideal points \don.
In particular, the ideal points added as endpoints to causal curves may 
not correspond to endpoints of spacelike curves, and vice-versa.
Discussion of these issues and corresponding improvement of the GKP
scheme is to appear in \mrtwo.
In what follows, we shall 
ignore this subtlety, and instead concentrate on identifications 
between the ideal points ensuing from the requirement of causality. 
This  allows us to talk only of the causal boundary of the 
spacetime, not the conformal boundary as one is usually accustomed to.

\newsec{Causal structure of generic plane wave spacetimes}

We now proceed to use the Geroch--Kronheimer--Penrose \geroch\ method
to ascertain the causal structure properties of plane wave backgrounds. 
The essential ingredients for the construction as we have discussed earlier 
are the knowledge of the TIP(F)s in the spacetime, and the identifications 
between them. 
This was done by \maro\ for the homogeneous plane waves \homplane\ in
detail and discussed for a few more general cases.
Here we review the method used in \maro\ (with some modifications), and
then extend it to find the causal structure of more general plane waves.

The generic plane wave has an arbitrary functional dependence 
on the coordinate $u$, through the function matrix $f_{ij}(u)$ appearing in 
\plane. One would presume that, without knowledge about some characteristic 
features of the matrix $f_{ij}(u)$, making detailed statements regarding the 
causal structure of the spacetime would be well nigh impossible. However, 
one can extract a lot of information about the causal structure without 
knowing the details of the functional form, simply by resorting to local 
analysis. In particular, we will be able to use the fact that $f_{ij}(u)$ 
is a real, symmetric matrix, with its elements being real continuous 
functions, to first approximate the functional form of $f_{ij}(u)$ by constants
$f_{ij}(u_0)$ in a neighbourhood of $ u = u_0$, implying that in this 
neighbourhood the metric is of the homogeneous plane wave form \homplane.
Having done so, we can perform a rotation in the transverse space to 
put the metric in the form
\eqn\diaplane{
ds^2 = -2 \, du \, dv - f_{i}^{(0)} \, (x^i)^2 \, du^2 + dx^i \, dx^i}
with $f_i^{(0)}$ being the eigenvalues of the matrix $f_{ij}(u_0)$. 
The only requirement we will have,
stemming from the energy conditions,
 is that ${\rm Tr}(f_{ij}(u)) \ge 0 $ for all $u$.

In the next subsection, we turn to the question of constructing the TIPs 
for a general plane-wave.  In terms of the coordinates used in \plane,
a future-endless curve can either end at infinite $u$ or at finite $u$
(in which case some other coordinate diverges).
We first turn to this latter case.

\subsec{Review of TIPs of the general plane wave}

We will now rephrase the proof, used by Marolf and Ross in the 
Appendix of \maro, of the  claim that 
the TIP of any causal curve $\g$ which asymptotes in
the future to a finite $u = u_1$ is given by the set of all points with
$u \le u_1$.  We write this claim in a somewhat condensed notation as
\eqn\claimA{
{\rm TIP}\[\g(u \to u_1)\] 
= \{(u,v,x^i): u \le u_1 \}}
Since $u$ is just a parameter along the curve, we can translate 
it by $u \to u + u_1$, so that \claimA\ is equivalent to the claim that
${\rm TIP}\[\g(u \to 0)\] = \{(u,v,x^i): u \le 0 \}$.
To prove this, it in fact suffices to show that
the TIP of any curve $\g$ asymptoting to $u=0$
 contains all points on the surface $u = -\d$ for arbitrarily small $\d$,
\eqn\claimB{
\forall \ \d > 0, \ \ \ 
{\rm TIP}\[\g(u \to 0)\] \supset \{(u,v,x^i): u = -\d \} }

The rest follows by the following series of steps:

\itm Any point with $u < -\d$ is in the past of some point in the
 $u = -\d$ surface, so \claimB\ implies that 
 $\forall \ \d > 0, \ 
 {\rm TIP}\[\g(u \to 0)\] \supset \{(u,v,x^i): u \le -\d \}$.

\itm No point with $u > 0$ can be in the TIP of any causal curve
 $\g$ which asymptotes to $u=0$, because $u$ can only increase
 along (future-directed) causal curves. 
 This can be seen as follows: 
Take any causal curve $\g$ and any point $p_0$ on $\g$.
We can use the plane symmetry to translate the $x^i$s to the origin
along the constant $u$ plane.  Now at $p_0$, the causal relation
(in the translated coordinates) then becomes
$-2 \, \ud \, \vd + (\xid)^2 \le 0$, which is exactly the same 
relation we would obtain in flat space.
But in the flat space, it is clear that $u$ cannot decrease along
any causal curve (since if it did, so would $v$ in order to maintain
causality, but then the curve would by definition be past-directed).
This shows that $\ud \ge 0$ at $p_0$, but since $p_0$ was arbitrary,
we have shown that $u$ can only increase
 along future-directed causal curves. 

\itm Finally, TIP is by definition a closed set, 
 so that taking $\d \to 0$ leads to the 
 desired statement that 
 ${\rm TIP}\[\g(u \to 0)\] = \{(u,v,x^i): u \le 0 \}$.

\noindent
Thus, if we can prove the claim \claimB, we are done.

To simplify the proof even further,
 we now make use of the transverse-plane symmetry of plane waves (which
is not present for generic pp-waves):  
we can translate the point $(u\!=\!-\d,v,x^i)$ 
along the wavefront to  $\ptd$.
We would of course need to apply the same translation to our
``set of all causal curves $\g(u \to 0)$'', but this set
 is by definition invariant.
Thus, to prove \claimB, it suffices to show that 
the point $\ptd$ is in the past of {\it any} causal curve which asymptotes 
to $u=0$,
\eqn\claimC{
\forall \ \d > 0, \ \ \ 
\CI^-\[\g(u \to 0)\] \ni  \ptd }
But by definition of 
$\CI^{\pm}$, 
for any two points 
$p$ and $q$, $q \in \CI^-(p)$ iff $p \in \CI^+(q)$.
This means that \claimC\ is equivalent to the claim
\eqn\claimD{
\forall \ \d > 0, \ \ \exists \  \e > 0 {\rm \ \ s.t.\ \ \ }
\g(u > -\e) \subset \CI^+\ptd }
\ie, any causal curve $\g$ which asymptotes to $u=0$
 must enter the future of the point $\ptd$.

Now, to show this, it of course suffices to prove that any causal 
$\g(u \to 0)$ enters into any subset of $\CI^+\ptd$;
in particular that it enters into the region bounded by null
(and therefore causal) curves $\CC$
emerging from $\ptd$.
Note that if these null curves $\CC$ are actually null
geodesics, this region is the full\foot{
This is actually true only locally, upto where the geodesics caustic.} 
$\CI^+\ptd$;
however, it may turn out more convenient not to require that
$\CC$ be geodesics.  In fact, \maro\ make such a non-geodesic
choice to complete their proof.

Specifically, \maro\ first choose convenient null curves $\CC$
(see their eqn.(A.1))
and  construct the region 
$R_{\d} \subset \CI^+\ptd$ bounded by these curves.
Far along the curves, this is given by 
\eqn\want{
x^2 \le 2  (1-\e_1) \, \d \, v}
for some $\e_1$, where $x^2 \equiv \sum_i x^i \, x^i$.
Thus, to prove \claimD, we have to show that
 the coordinates along any causal curve $\g$ 
satisfy the relation \want\  sufficiently far along $\g$ (\ie, 
for $u$ sufficiently close to zero). Recall that for any future-endless
$\g$ asymptoting to a finite $u$ plane, at least one other coordinate must 
diverge along $\g$; in fact, $v$ must diverge, since by causality $v$ grows 
faster than any $x^i$.
Now, if $v \to \infty$ as $u \to 0$ while $x$ stays bounded, 
equation \want\ is automatically satisfied, so we are done.
Therefore we will assume that $x$ also diverges as $u \to 0$,
\ie, $\g$ reaches arbitrarily large values of $x$ and $v$.
To bound how fast $x$ can diverge relative to $v$,
 \maro\ use the metric to argue that as $u \to 0$, 
\eqn\causal{
2 \vd \ge (1-\e_2) \, \xd^2 }
along any causal curve $\g$ for arbitrarily small $\e_2$.
The causal relation \causal\ can be reproduced by a fiducial metric
\eqn\fiducial{
ds^2_{\rm fid} = 
 -2 \, du \, dv  + (1-\e_2) \, dx^i \, dx^i}
which gives the finite-difference relation\foot{
This works because \fiducial\ is flat---otherwise
we would have to integrate $ds$ to obtain the correct $\Delta s^2$.}
$(\Delta x)^2 \le {2 \over 1-\e_2} \, \Delta v \, \Delta u$.
Since $x,v \to \infty$, $\Delta x \to x$ and $\Delta v \to v$, 
so the relation becomes $x^2 \le {2 \over 1-\e_2} \, v \, \Delta u$.
But for $\Delta u \equiv \e \ll \d$, we can easily arrange for
$x^2 \le 2  (1-\e_1) \, \d \, v$, which is exactly the relation
\want\ describing the set contained in $\CI^+\ptd$.
Hence, given any (arbitrarily small) $\d$,
we have shown that for $u$ sufficiently close to $0$,
$\g(u \to 0)$ enters $\CI^+\ptd$. \ 
{\bf (QED)}

\subsec{Comments on nonexistence of horizons}

Let us at this point take a detour from constructing the full causal 
structure of plane waves, and comment on the implications  of the 
results thus far, concerning the nonexistence of event horizons as
discussed in \nobh.
In particular, knowledge of the TIPs will immediately tell us that 
the spacetimes which are of the plane wave form  can not admit 
event horizons. 

Recall that a necessary condition for the 
existence of an event horizon is that there exist points in the
spacetime which are causally disconnected from future infinity,
$\Ci^+ \cup \scri^+$.
Hence, to prove the absence of horizons, it suffices to prove that
all points in the spacetime are contained in the past of infinity.
By proving that  for any finite value of $u_1$,
all points with $u \le u_1$ are in the past of the ideal point
corresponding to the TIP of any causal curve $\g$ which asymptotes
to $u=u_1$, we have shown that all of spacetime is causally connected
to infinity.
This in turn means that there can't be any black holes of the 
plane wave form, \ie, that plane waves don't admit event horizons.

There is however one subtlety, which we now address.
The argument of \maro\ applies to general plane wave metrics \plane,
with the assumption that $f_{ij}(u)$ is regular in open neighbourhood 
of $u= u_1$.
Now, since $f_{ij}(u)$ is not required to remain regular in a general
plane wave, let us classify plane waves into two classes:  
(1) nonsingular, with 
$f_{ij}(u)$ remaining finite for all values of $u$, and
(2) singular, where
some $f_{ij}(u) \to \infty$ as $u \to u_{\infty} < \infty$.
For nonsingular plane waves, we can
consider any point $p_0=(u_0,v_0,x_0^i)$ of the spacetime, and
 apply the above proof for $u_1>u_0$, to show that 
$p_0 \in {\rm TIP}\[\g(u \to u_1)\]$, or $p_0 \in \CI^-\[\scri^+\]$.
For the second class, with the spacetime becoming singular at some
finite value of $u \to u_{\infty}$, we can apply essentially the 
same argument as above: 
The spacetime manifold is an open set (which in particular does not
by definition include the singularities), so for any point 
 $p_0=(u_0,v_0,x_0^i)$ in the spacetime, there exists $u_1$ such that
$u_0 < u_1 < u_{\infty}$.  Applying the proof of \maro\ to $u_1$,
we again see that $p_0 \in {\rm TIP}\[\g(u \to u_1)\]$, \ie, $p_0$ 
is visible from infinity.

\subsec{Causal structure of plane waves}

Let us now return to the more general discussion of the causal 
structure of plane waves.
The construction outlined in  subsection 4.1,
in particular the claim \claimA, tells us that 
the TIPs are parameterised by a single parameter $u$.\foot{
So far, we have said nothing about the TIPs of causal curves 
which reach infinite $u$.  However, since this is closely tied to 
the  discussion of identifications, we shall first address those, and
return to this point afterwards.}
By exchanging past and future, the same conclusion will apply 
to the TIFs.
To complete the discussion of 
the causal structure, we need to know what are all the  identifications 
between the TIPs and the TIFs.
We will first review the simple case of maximally symmetric homogeneous 
plane wave addressed by \maro, and then extend this to more general 
plane waves.

We will start with the simplest case of the maximally supersymmetric plane 
wave solution to Type IIB supergravity, given by the metric \plane\
with $f_{ij}(u) = \mu^2 \delta_{ij}$. It is shown in \maro\ that 
for every $u_0$, the TIP described by  
$\{(u,v,x^i): u \le u_0 \}$ is identified 
with the TIF $\{(u,v,x^i): u \ge u_0   + {\pi \over \mu} \}$. 
In other words, the ideal points  corresponding to, heuristically,
$(u=u_0, v \to + \infty, x^i)$ and 
$(u=u_0+{\pi \over \mu}, v \to - \infty, x^i)$ are identified.
This identification 
rests on the simple fact that the future of every point on the 
$u = u_0$ plane contains the future of all points on the 
$u = u_0 + {\pi \over \mu}$ plane. To see this, it suffices to show that 
there exists a sequence of null geodesics emanating from any point on
the $u = u_0$ plane, {\it i.e.}, null geodesics starting from 
$p_0 = (u_0, v,x^i)$ for arbitrary $v,x^i$, which have an accumulation curve 
ending up at $p_1 = (u_0 + {\pi \over \mu}, - \infty, x^i)$ for some finite 
but arbitrary $x^i$.  
Furthermore, the same is not true for any smaller value of $u$ in $p_1$.
The existence of this accumulation curve for the 
sequence of null geodesics implies that we can causally reach 
large negative values of $v$ in finite $u$. By 
construction of the sequence, we know that $p_1$ is in the causal future of 
$p_0$, so all points with $u \ge u_0 + {\pi \over \mu}$ are in the 
causal future of $p_0$ as well. This is the reason why we need to identify 
the TIPs and the TIFs: the associated ideal points share the same 
causal past and future. 

This identification has a very important consequence.
In particular, taking $u_1 \to \infty$ in the above argument, 
we see that
any causal curve which reaches infinite $u$ has in its past the
entire spacetime, so that the whole region of the causal boundary
corresponding to infinite $u$ is actually represented by a single
ideal point, $\Ci^+$.
Thus, having identified the TIPs and the TIFs we see that the causal 
boundary of the spacetime is parameterized by a single parameter $u$, 
\ie, it is described by a single line!
As we mentioned in Section 3, the construction of the ideal points allows us 
to discuss the causal properties of the boundary in the completed spacetime.
In particular, as was already demonstrated by \maro, this curve of ideal
points is locally null.\foot{
Since the identifications cause this line boundary to ``wind around''
(as can be easily checked by conformally mapping the spacetime into
the ESU and seeing the corresponding boundary wind around the cylinder),
the ideal points separated by $u > {\pi \over \mu}$ are in fact
timelike-separated.}

\ifig\figBdy{Causal boundary of a plane wave, $\scri$: 
In most cases it is a one-dimensional null line, 
with some TIPs and TIFs identified, as
exemplified by the causal curves $g1$ and $g2$.}
{\epsfxsize=4cm \epsfysize=8cm \epsfbox{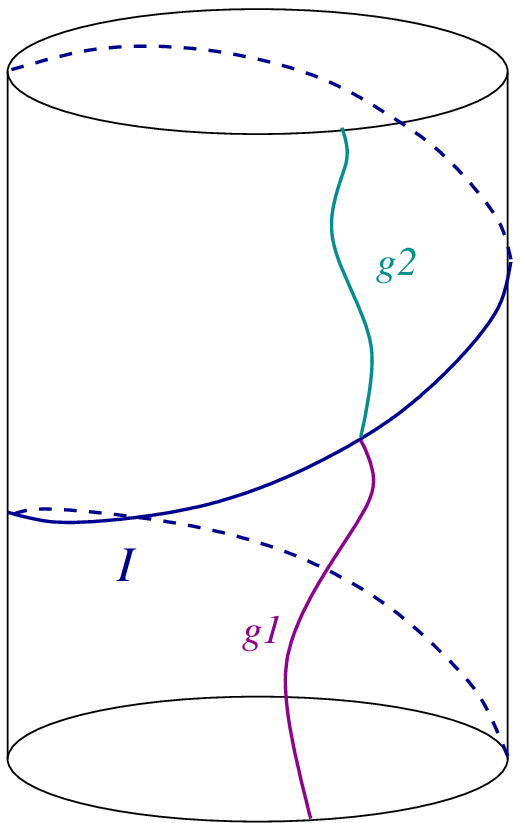}}

This is the first example, promised above, of a null boundary which 
nevertheless has its ideal points corresponding to both TIPs and TIFs.
This point is illustrated in Fig.2.
The spacetime is conformally mapped to the cylinder 
(in this particular case,  representing the Einstein Static Universe), 
and its causal boundary $\scri$ then winds
around this cylinder in a null fashion as shown.  Causal curves, 
such as  $g1$, can end on this boundary; but they can also 
 start on it, as $g2$ in Fig.2.
Note, however, that geodesic observers cannot ``pass through'' this 
boundary, because it takes them an infinite proper time to reach it,
as guaranteed by the fact that the boundary is non-singular.  (This will 
be contrasted later, for a certain class of pp-waves.)
As we will see below, this ``null line'' nature of the boundary 
appears much more generically than just for this conformally flat 
homogeneous plane wave.

In the case of the generic plane wave, determination of the 
identifications between the TIPs and TIFs is generally more complicated, 
since the geodesic equations depend now on the explicit form of the 
functions $f_{ij}(u)$. 
Specifically, as derived in Appendix A, the geodesic equations are
given by 
\eqn\geodx{
\xdd^i + \sum_j f_{ij}(u) \, x^j  = 0}
and $v(u)$ is determined simply from integrating
the first order constraint equation,
\eqn\veq{
v = {1 \over 2} \, \sum_i x^i \, \xid + v_0}
where we can take $\dot{} \equiv {d \over du}$, and
$v_0$ is an arbitrary integration constant which is fixed by the 
initial conditions.

To see if there are any identifications and to claim 
that the causal boundary is one-dimensional, it however suffices to 
show that, starting from any $p_0 = (u_0, v,x^i)$,
 there is a sequence of null geodesics which accumulate toward 
large negative values of $v$ whilst keeping the transverse coordinates 
finite. This will in particular happen so long as the behaviour 
of the null geodesics continues to be oscillatory as a function of 
$u$, for arbitrarily large values of $u$.  Here we are assuming that the 
functions $f_{ij}(u)$ are regular functions of $u$. In the case that they are
not, we will encounter singularities at the null planes where $f_{ij}(u)$ 
diverges, as discussed above. 

Before we embark on explicitly demonstrating that this is possible for a wide 
variety of examples, let us for a moment pause to look at the 
situation in the BMN plane wave, with $f_{ij}(u) = \mu^2 \delta_{ij}$. 
Here, the geodesic equations \geodx, \veq\ can be solved to give
\eqn\bmnnullg{\eqalign{
x^i(u) & = {\sqrt{2 \kappa} \over \mu} \sin \( \mu u \) \cr
v(u) & = {\kappa \over 2 \mu} \sin \(2 \mu u\)
}}
where we have chosen to look at null geodesics emanating from 
the origin, with velocity $\dot{x}^i(0) = \sqrt{2 \kappa} \; \forall \; i$.
In the neighbourhood of $u \to \({\pi \over \mu}\)^-$ we 
see that $x^i(u) \rightarrow 0^+$ with $\dot{x}^i(u) <0$, implying 
$v(u) < 0$. Now with a suitable choice of the parameter 
$\kappa$ one can ensure that $v(u \to {\pi \over \mu}) \rightarrow - \infty$, 
whilst retaining finite values of $x^i(u={\pi \over \mu})$. 
For instance, consider a sequence of null geodesics 
labeled by $\kappa_n = {1\over 2 } n^{3/2}$. Considering the 
sequence of points on the aforementioned  null geodesics at 
$u_n = {\pi \over \mu} -{1 \over n}$, in the limit $n \rightarrow 
\infty$, we find that these have an accumulation point,  
$(u = {\pi \over \mu},  v = - \infty, x^i =0)$.
Thus, for the  
existence of  a sequence of null geodesics that accumulate towards large 
negative values of $v$, with finite values of transverse coordinates, it 
suffices to show that the geodesic equations \geodx\ admit 
solutions wherein the curves $x^i(u)$ have zeros at arbitrarily 
large values of $u$, with a negative slope. Given that $v$ scales 
quadratically with the transverse coordinates, it is possible to 
choose initial conditions such that $v$ gets arbitrarily large and 
negative. If this behaviour persists for arbitrarily large values of $u$, 
then the identifications between the TIPs and the TIFs ensure that the 
causal boundary is one-dimensional.

It is easy to see that this accumulation point criterion will be satisfied
for $f_{ij}(u)$ which are polynomials, trigonometric, or hyperbolic functions. 
It is possible to make explicit statements if we assume that the matrix 
$f_{ij}(u)$ is of the form $f_{x j} (u) = 0$ for $j \ne x$ and let
$f_{xx}(u) = f(u)$  for some transverse space coordinate $x$. This 
simplification will allow us to analyze the geodesic equation \geodx\ without 
concerning ourselves with having to solve a system of coupled oscillators. 
Let us now try to understand the behaviour of the geodesics for 
particular cases of the function $f(u)$. Our notation of 
special functions conforms to the standards of \gradr.

\vskip3mm
\noindent
{\bf 1.} $f(u)$ is a polynomial function of $u$, say $f(u) = u^n$. 
In this case we see that the solution to the geodesic equation 
$ \ddot{x}(u) + u^n \, x(u) =0$ is given in terms of Bessel functions, 
$x(u) = \sqrt{u} \, J_\nu({2 \over n+2} \, u^{n+2 \over 2} )$, with 
$\nu = \pm {1 \over 2 +n}$. These solutions are clearly oscillatory for 
arbitrarily large $u$, given that $\sqrt{u} \, J_\nu(u) \rightarrow 
\cos(u -{\pi \over 2} \nu - {\pi \over 4})$ for $ u \gg 1$. 

\vskip3mm
\noindent
{\bf 2.}
 $f(u)$ is an exponential function, say $f(u) = e^u$. Again the solutions 
are Bessel functions with $x(u) = J_0(2 e^{{u \over 2}})$ or 
$x(u) = N_0(2 e^{{u \over 2}})$, which exhibit oscillatory behaviour 
for arbitrarily large $u$.

\vskip3mm
\noindent
{\bf 3.}
 Consider $f(u) = \cos(u)$. The solutions to \geodx\ are 
$x(u) =  {\rm ce}_0(x,-2)$ or $x(u) = {\rm se}_0(x,-2)$, where
${\rm ce}_{a}(x,q)$ and ${\rm se}_a(x,q)$ are the periodic 
Mathieu functions. In this case also we have an oscillatory behaviour.

\vskip3mm
\noindent
{\bf 4.}
 An interesting example to consider is one where $f(u) \rightarrow 0$ 
for large $u$. One would imagine that in this case, with 
rapid approach to flat space, the status of the identifications between the 
TIPs and the TIFs would be problematic. Let us for concreteness 
consider $f(u) = {1 \over 1+ u^2}$. For this example we can show that 
$x(u) = {\rm F} \( - {1\over 2} (-1)^{1/3}, {1\over 2} (-1)^{2/3}; 
{1 \over 2}; -u^2 \)$ or $x(u) = u \, {\rm F} 
\( {1\over2} - {1\over2}(-1)^{1/3}, 
{1\over2}+ {1\over2}(-1)^{2/3} ; {3 \over 2}; -u^2 \)$, where 
${\rm F}(a,b;c;x)$  denotes the hypergeometric function. 
Here too it is easy given the explicit solution to convince oneself 
that there exist identifications.

\vskip3mm
\noindent
{\bf 5.}
Suppose, on the other hand the approach to flat space is exponential, \ie,
$f(u) = e^{-u^2}$. In this case the geodesics are oscillatory in a 
small neighbourhood of $u =0$. This oscillatory behaviour ceases for some 
finite $u$. While there exist identifications for finite values of $u$,
 the structure at large $u$ is akin to flat space. Thus, the causal 
structure of this spacetime is similar to the sandwich plane wave as 
discussed in \maro.

\vskip3mm
\noindent
{\bf 6.}
 Let us turn to a slightly different behaviour of $f(u)$, one wherein 
we have a singularity. Consider for instance, $f(u) = {A \over u^2}$. 
This is the generic behaviour of plane waves limits of black hole
spacetimes, when we consider the null geodesics which terminate at the 
singularity. For this form of $f(u)$, it is easy to see that there 
are oscillatory solutions as long as $A > {1\over 4}$.
\vskip3mm

In the preceding discussion we have investigated certain generic classes 
of functional behaviour for which it is possible to show explicitly where 
there are identifications between TIPs and TIFs.
Let us now see what happens for 
plane wave spacetimes which can be obtained as Penrose limits 
of some geometries which are interesting from a string theory point of view. 
Certain Penrose limits of D-branes were discussed in \maro;
here we propose to consider the plane wave limits of 
spacetimes which holographically encode dynamics of non-local theories,
such as little string theory or non-commutative gauge theory. The 
Penrose limits for these spacetimes were discussed in \nonlocal. 

\vskip3mm
\noindent
{\bf NL1.}
 The Nappi-Witten geometry \napwit,
\eqn\nwit{
ds^2 = -2 \, du \, dv - \l^2 \, \vec{z}^2 \, du^2 + d\vec{z}^2 
+ \sum_{i=1}^6
\,  (dy^i)^2
}
is obtained as the Penrose limit of the near horizon geometry of NS5-branes
\googuri,\kirit, \nonlocal, by considering null geodesics which have 
some angular momentum $\l$ 
along the $\S^3$ transverse to the NS5-brane.
In the geometry \nwit, we have two directions labeled by the vector $\vec{z}$, 
for which $f_{\vec{z}}(u) = \l^2$. In this case, there are clearly 
oscillatory null geodesics. This situation is no different from that 
in the BMN plane wave. Here we will have identifications between the 
TIPs and the TIFs 
for arbitrarily large values of the coordinate $u$ and in particular, 
we can conclude that the causal boundary in this case is one-dimensional.
This is perhaps a little surprising, as one might have expected that the 
geometry is a direct product of a four dimensional plane wave of the BMN kind 
(in the sense that two directions have positive mass terms on the world-sheet
in the light-cone gauge), with a six dimensional flat space. However, with the 
persistence of the identifications between the ideal points for the TIPs and
the TIFs for arbitratily large values of $u$, the entire spacetime 
is visible from $u = \infty$, and hence the causal boundary collapses to 
a one-dimensional null line. 

\vskip3mm
\noindent
{\bf NL2.}
 We can consider the Penrose limit of the near-horizon geometry of 
NS5-branes, by looking at null geodesics which in addition to 
angular momentum on the transverse $\S^3$ also have a radial component.
The plane wave geometry resulting from the Einstein frame metric of the 
NS5-brane is \nonlocal,
\eqn\einsnsfive{
ds_{E}^2 = - 2 \, du \, dv - {1 \over 4u^2} \(\vz^2 + x^2 + 
\sum_{i=1}^5 \, y_i^2 + b  \, u \, \vz^2 \) du^2
+ dx^2 + d\vz^2 + \sum_{i=1}^5 \, dy_i^2, 
}
with $b = 8 {\l^2 \over \sqrt{1-\l^2}} > 0 $. Clearly, the 
geometry has a null singularity at $u =0$. 
For the six directions labelled by $(x,y_i)$, 
the matrix $f_{i j}(u)$ is diagonal with the entries being ${1 \over 4 u^2}$.
As discussed in the example 6 above, in this case the null geodesics 
along none of these directions exhibit oscillatory behaviour.
Along $\vz$ however, it is easy to see that there are oscillatory solutions. 
Writing $\vz = (z^1,z^2)$ we see that the general 
solution for the 
geodesic equation is $z^i(u) = \a^i_1 \, \sqrt{u} \, J_0(2 \, \sqrt{b} \, u)
+ \a^i_2  \, \sqrt{u} \, Y_0(2 \, \sqrt{b} \, u)$, where 
$\a^i_1$ and $\a^i_2$ are 
arbitrary constants. These geodesics exhibit oscillatory behaviour 
for arbitrarily large values of the $u$ coordinate.

An interesting feature that can be explicitly discussed with example NL2,
pertains to the distinction between the string and Einstein frame metrics 
with regard to the causal structure. Since the string frame metric is 
related to the Einstein frame metric by a conformal transformation, one 
expects that for dilaton profiles which are non-singular, the discussion 
with respect to the causal structure to be identical. In the case of 
singular behaviour of the dilaton we would encounter additional 
singularities in the Einstein frame metric. The point that we wish to clarify
 is that while taking the Penrose limit, one can work either with the string 
frame or the Einstein frame metric, and the resulting plane wave spacetimes 
are again related to each other by a conformal rescaling. So unless the 
Penrose limit induces additional singularities in the dilaton profile, the 
causal structure for both is identical. In the case of the NS5-brane 
geometry, if we had taken the Penrose limit for the string frame metric, we 
would have obtained the plane wave metric and a dilaton, 
\eqn\strnsfive{\eqalign{
ds_{str}^2 &= - 2 \, du \, dv -
\l^2 \, \vz^2 \, du^2
+ dx^2 + d\vz^2 + \sum_{i=1}^5 \, dy_i^2  \cr
\phi(u) & = - \sqrt{1 - \l^2} \, u.
}}
The string frame metric is identical to the Nappi-Witten model \nwit, 
and one concludes that the geodesics in the $\vz$ directions lead to 
identifications. The dilaton being linear in $u$ leads to a singularity 
in the spacetime at $ u =0$. This is precisely what we conclude from the 
Einstein frame metric \einsnsfive.

\vskip3mm 
\noindent
{\bf NL3.}
 The last example we wish to consider is the Penrose limit 
of the geometry which is holographically related to the non-commutative 
Yang-Mills theory \akisunny, \mrusso.
In this case the choice of a null geodesic with angular and radial 
components leads to the 
 plane wave metric \nonlocal
\eqn\ncplane{
ds^2 = - 2 \, du \, dv 
- \l^2 \, \(x^2 + \vz^2 + y_1^2 + g(u) \, y_2^2\) \, du^2 
+ dx^2 + d\vz^2 + dy_1^2 + dy_2^2,
}
where $\l \le 1$ and 
\eqn\ncppg{
 g(u) \equiv {\l^8 - a^8 \sin^8(\l u) - 2 a^4 \sin^2(\l u) \, \cos^2(\l u)
\( a^4 \sin^4(\l u) - 5 \l^4 \) \over
\( \l^4 +  a^4 \sin^4(\l u) \)^2 }.
}
Here, the spacetime as a whole has identifications between the TIPs 
and the TIFs, because one has geodesics behaving like harmonic 
oscillators in all of the transverse directions but $y_2$, for $\l \neq0$.
However, it is possible for the identifications to disappear if we 
consider special limits of the parameters appearing in the metric \ncplane,
\ncppg. This rests on the behaviour of null geodesics along $y_2$, 
which are determined by the geodesic equation
 $\ddot{y}_2(u) +  \l^2 \, g(u) \, y_2(u)=0$. 
While it is hard to explicitly analyze the geodesic equation 
analytically, one can numerically integrate it to show that generically 
there are oscillatory solutions. The situation changes when we consider
the strong non-commutativity limit $a \rightarrow \infty$ and $\l =1$. 
In this case $g (u) \rightarrow \,
-  \(1 + 2 \cot^2 u\)$. This implies that the spacetime is singular at 
$u=0,\pi$ and so we take the coordinate $u \in (0,\pi)$. The geodesics 
along the other transverse directions $(x,\vz,y_1)$ are oscillatory 
with period $2\pi$ (since we take $\l =1$). Thus, the identifications 
between TIPs and TIFs disappear, since they 
would require a larger separation of the $u$ coordinate 
than the allowed range $u \in (0,\pi)$.

As we have seen above and as 
was pointed out already in \maro, there exist examples of spacetimes wherein 
there is no identification, such as the plane waves arising from the 
Penrose limit of D0, D1, and D2 brane near horizon geometries.
In these cases, the causal boundary comprises of a null line for finite
values of $u$, and past and future null planes (rather than single ideal 
points) corresponding to $u \to \pm \infty$. This obstructs the statement 
that the full causal boundary is one-dimensional. 
More trivially, the flat Minkowski spacetime has no TIP $ \leftrightarrow$ TIF
identifications, and higher dimensional causal boundary.
Since this may appear somewhat puzzling, let us make a few remarks 
about the distinctions between these plane waves, before 
proceeding to discuss the more general pp-waves. 
For simplicity, let us illustrate the point by discussing the
Minkowski spacetime example.


\subsec{Why is flat spacetime different?}

Above, we have seen that if the geodesics have an oscillatory 
behaviour in $v$ (as a function of $u$), we can find appropriate
sequences which have an accumulation point at large negative 
values of $v$.  This means that for some $u_0$ and $u_1$,
all points with $u \ge u_1$ lie in the future of all points with
$u \le u_0$, so that the ideal points corresponding to the TIP
$(u_0, v \to + \infty, x^i)$ and the TIF
$(u_1, v \to - \infty, x^i)$ get identified.

This identification has a dual role.  
Firstly, it tells us that instead of two lines of ideal points,
one corresponding to TIPs and the other to TIFs of causal curves
asymptoting to a finite value of $u$, we have only a single line,
which ``winds around'' as in Fig.2.
More importantly, it also tells us that the TIPs of curves asymptoting
to an infinite $u$ all give a single ideal point $\Ci^+$, and similarly,
the TIFs of all curves starting from negative infinite $u$ all give rise
to a single ideal point $\Ci^-$.  This completes the statement
that the causal boundary of such a spacetime is one-dimensional.

Let us now contrast this situation with that of the flat 
$d$-dimensional spacetime, which is a special class of plane waves.
Since the claim of Section 4.1 that 
\eqn\claimAr{
{\rm TIP}\[\g(u \to u_1)\] 
= \{(u,v,x^i): u \le u_1 \}}
holds for all plane waves (and therefore the flat spacetime), 
we might deduce that the boundary is parameterized by a single parameter
$u$ and thus is one-dimensional.  However, this clearly contradicts the
obvious fact that a $d$-dimensional Minkowski spacetime
has a $(d-1)$-dimensional boundary.  
What went wrong?
The flaw in the above reasoning stems from the lack of identifications.
Since the geodesics in flat space are just straight lines, they can't
reach arbitrarily large negative $v$.  
In fact, in flat spacetime, $v$ must increase along all 
future-directed causal curves.
The set of ideal points
$\{ {\rm TIP}\[\g(u \to \infty)\] \}$ corresponding to infinite $u$
is not zero-dimensional as above, but rather a full $(d-1)$-dimensional
surface.
Put more explicitly, most curves which reach infinite $u$, unless they
simultaneously reach the future timelike infinity $\Ci^+$,
do {\it not} have the whole spacetime in their
past. This is illustrated in Fig.3.

\ifig\figMink{Penrose diagram for Minkowski spacetime.
The TIP of a causal curve $g1$ which asymptotes to the $u=0$ plane
lies in the null cone $u = 0$ shown.  
Since the part of $\scri^+$ corresponding to finite $u$ is the 
upper left line indicated,
the curve $g2$ does not asymptote to any finite $u$.}
{\epsfxsize=9cm \epsfysize=9cm \epsfbox{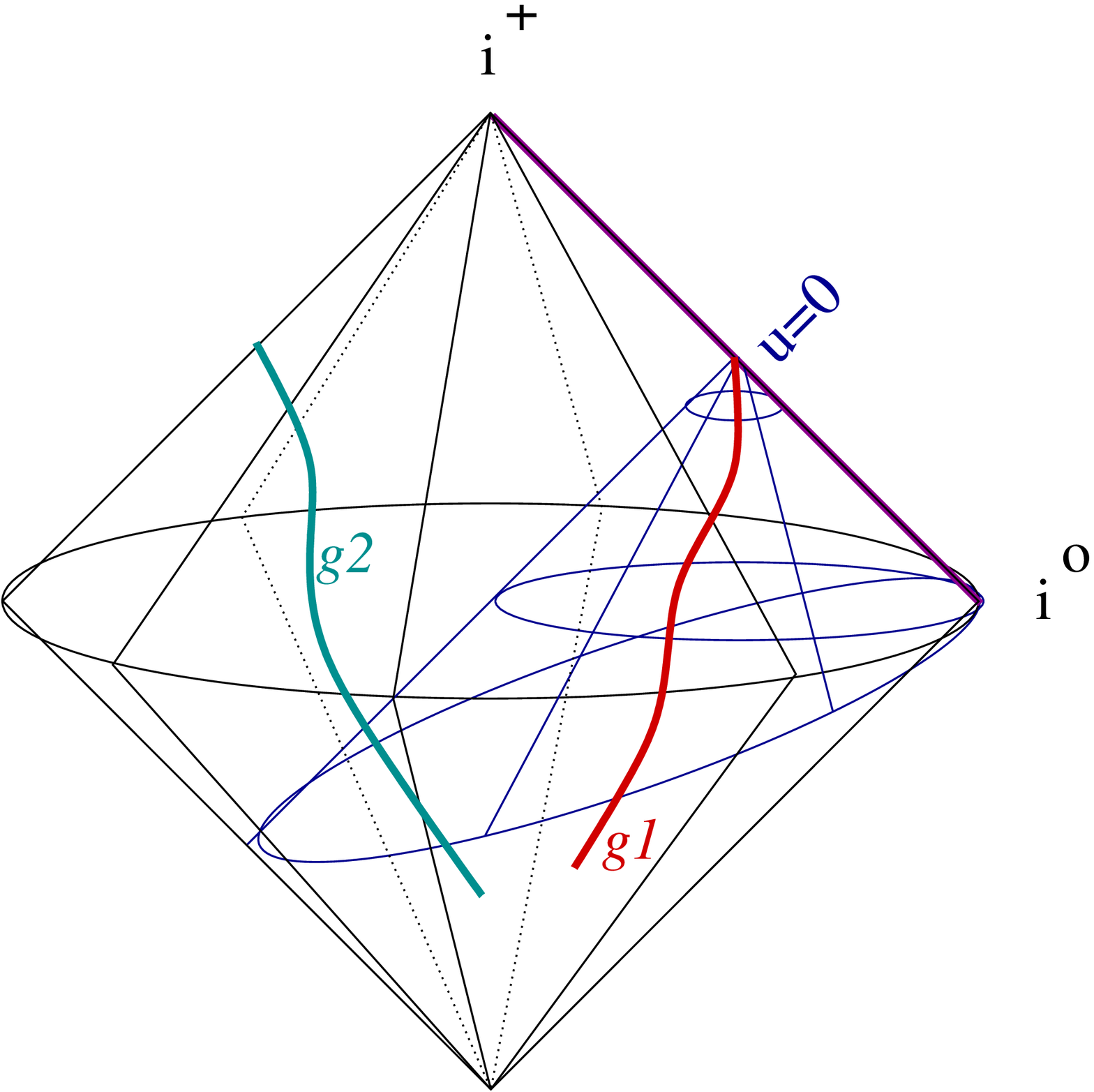}}

Fig.3 shows the Penrose diagram for the flat Minkowski spacetime,
with $(u,v,r)$ directions shown explicitly.  The whole spacetime 
is bounded by 2 null cones (the upper corresponding to $\scri^+$ and
the lower to $\scri^-$; in addition there is spatial infinity $\Ci^0$ and
the future/past timelike infinities $\Ci^{\pm}$ 
lying at the tips of the cones).
 The constant $u$ or $v$ null planes 
are here conformally mapped to null cones inside the spacetime.  
For example, the $u=0$
hypersurface is the null cone shown in Fig.3.
Therefore, the part of $\scri^+$ corresponding to {\it finite}
$u$ is the null {\it line} in the upper right edge of the diagram; the
rest of $\scri^+$ has infinite $u$.  This is what accounts for the
higher dimensionality of the boundary.
The claim ${\rm TIP}\[\g(u \to 0)\] = \{(u,v,x^i): u \le 0 \}$
is now obvious from Fig.3 (where $\g$ is exemplified by the curve $g1$).
However, we can also easily see that
there are now curves, such as $g2$, which reach infinite $u$ without
reaching $\Ci^+$, and therefore do not contain all of the spacetime
in their past.

This property of flat space is also what makes it the only exception
to the observation \penr\ that plane waves are not 
globally-hyperbolic.  In particular, for all non-trivial plane waves,
the light cones eventually re-converge in caustics, so that there 
exists a sequence of null geodesics converging to two distinct, parallel,
null geodesics.  This property precludes the existence of global
Cauchy surfaces. 
Taking the flat space limit effectively pushes these two null geodesics
infinitely far apart; or in other words, the null geodesics do not caustic.

\newsec{Comments on causal structure of pp-waves}

We now turn to discussing the causal structure of general pp-waves,
with metric as given in \pp. 
In order to simplify the discussion 
we will write this metric in spherically symmetric coordinates, 
\eqn\pparad{
 ds^2 = -2 \, du \, dv - F(u, r, \Omega) \, du^2 + dr^2 + r^2\, d\Omega^2}
where, for simplicity of notation, we will refrain from writing out all the 
angles and instead content ourselves to the minimal symbol of 
$\Omega$.
The main advantage of this coordinate system is that along 
any causal curve $\g(u \to u_1)$ asymptoting to a finite value of $u$,
only two coordinates, $r$ and $v$, can diverge.
Furthermore, we can show that $v$ must diverge faster than $r$, so that
along any such causal curve $v$ necessarily diverges.

Let us first consider the class of pp-waves which are
solutions to vacuum Einstein's equations.
Since the Einstein tensor is given by
$G_{uu} = {1 \over 2} \nabla_T^2 F$, where $\nabla_T^2$ is the
transverse Laplacian,
 $F(u,r,\Om)$ of \pparad\
must satisfy the transverse Laplace equation, $\nabla_T^2 F = 0$.
This is a very remarkable result, since it implies that,
due to the linearity of Laplace equation, 
we may superpose the solutions.
In particular, we can decompose $F$ in terms of the 
$(d-3)$-dimensional spherical harmonics $Y_L(\Om)$, 
where $L \equiv \{ \l,m,... \}$:
\eqn\F{
F(u,r,\Om) = \sum_L \{ f_L^+(u) \, r^{\l} \, Y_L(\Om)
          +  f_L^-(u) \, r^{-(d-4+\l)} \, Y_L(\Om) \} }
In the neighbourhood of $u = u_0$ and $\Om = \Om_0$, where the functional 
behaviour of  $F(u,r,\Om)$ with respect to the $u$ coordinate is 
regular\foot{
Regularity of $F(u,r,\Om)$ with respect to the angular variables is 
obvious; 
and if  it so happens that the function $F(u,r,\Om)$ in \pparad\ has a 
singularity at finite $u$, say at $u = u_{\infty}$ then we must impose the 
additional restriction that $u_0 < u_{\infty}$.}, we can write
\eqn\Fapprox{
F(u_0,r,\Om_0) \equiv f(r) 
= \sum_{\l} \{ f_{\l}^+ \, r^{\l}  +  f_{\l}^- \, r^{-(d-4+\l)} \} }
As discussed below, there can be singularities at $r=0$ and/or
$r = \infty$.

For pp-wave spacetimes which are not 
solutions to vacuum Einstein's equations, 
the discussion in the following subsection will still carry through. 
At present, it is not completely clear as to what are all the 
pp-wave solutions, say to just supergravity equations of motion. 
We will therefore concentrate on some examples that are of 
interest to string theorists. The main example we consider is 
the solution written down in \malmaoz, a supersymmetric 
background of IIB string theory, which leads to the
$\CN =2$ sine-Gordon theory on the world-sheet in 
light-cone quantization.
 The metric for the solution is given by 
\eqn\mm{
ds^2 = -2 \, du \, dv - \( \cosh x - \cos y \)\, du^2 + dx^2 + dy^2 +
dz^i\, dz^i}
where we have reverted back to the Cartesian coordinates. 

In the following we will first discuss the construction of TIP(F)s for 
pp-waves \pparad. We show that as in the plane wave case, the TIP of 
all causal curves asymptoting to $u = u_1$ plane 
contains all points with $ u \le u_1$. Then we turn to the question of 
identifications between TIPs and TIFs. This will depend on the 
specific form of the metric, \ie, the choice of the function $F(u,r,\Om)$.
We will show that for the spacetime in \mm\ the TIPs and TIFs get 
identified as in the BMN plane wave case, and therefore the 
causal boundary is one-dimensional. On the other hand, for the 
vacuum pp-waves \F\ (which are not plane waves)
we will see that there are no identifications between the TIPs and TIFs. 
These spacetimes are also generically
geodesically incomplete in contrast to the spacetime \mm.

\subsec{TIPs for general pp-waves}

We will now proceed to analyze the TIP structure of  pp-waves which 
are of the form given in \pparad. Our main claim will be
analogous to \claimA,
\eqn\claimApp{
{\rm TIP}\[\g(u \to u_1)\] 
= \{(u,v,r,\Om): u \le u_1 \}}
or in words, the TIP of any causal curve $\g$ which asymptotes in
the future to $u = u_1$ is given by the set of all points with
$u \le u_1$. 
As in the plane wave case, if the function $F(u,r,\Om)$ is 
singular at some finite $u = u_{\infty}$, we will require that 
$u_1 < u_{\infty}$.

To establish our claim \claimApp, we will proceed in parallel with the 
situation in the case of plane waves. Again it will suffice for us to 
show that the TIP of any causal curve $\g$ asymptoting to $ u = u_1$ 
contains all points on the surface $ u = u_1 - \delta$ for 
arbitrarily small $\d$, \ie,
\eqn\claimBpp{
\forall \ \d > 0, \ \ \ 
{\rm TIP}\[\g(u \to u_1)\] \supset \{(u,v,r,\Om): u = u_1-\d \} }
Once \claimBpp\ is shown, 
the rest is again established by noting that 
all points with $u < u_1-\d$ are in the past of some point in the
 $u = u_1-\d$ surface, 
whereas no point with $u > u_1$ can lie in the 
${\rm TIP}\[\g(u \to u_1)\]$\foot{
Proving that $u$ cannot decrease along a future-directed causal curve 
here is slightly less trivial than the corresponding proof for plane waves,
since there we used the extra planar symmetry.
However, we can use a completely different argument, which is 
similar to that which we used in \nobh\ to argue for absence of horizons
in pp-waves:
In any finite, but arbitrarily large, region between the origin and infinity,
we can bound the function $F$ in \pparad\ from above, $F(u,r,\Om) \le F_0$.  
Then any curve which is causal in \pparad\ must be causal in a spacetime
of the form \pparad\ with $F(u,r,\Om)$ replaced by $F_0$.  But the latter
is just the flat spacetime, wherein we know that $\ud \ge 0$, 
since $u$ is not affected by the
coordinate transformation to explicitly flat spacetime.  This shows that
the same must hold in the present case.  An alternate proof is given 
in \eg\ \candela.
},
so that by closure, taking $\d \to 0$ leads to the 
 desired statement that 
 ${\rm TIP}\[\g(u \to u_1)\] = \{(u,v,r,\Om): u \le u_1 \}$.

To prove the claim \claimBpp, we note that it is equivalent
to the claim that for arbitrarily small positive $\d$,
the future of any point $p_0 = (u_0, v_0, r_0, \Om_0)$
with $u_0 \equiv u_1-\d$ necessarily contains a part of any 
causal curve which asymptotes to $u= u_1$.  In other words,
\eqn\claimDpp{
\forall \ \d > 0, \ \ \exists \  \e > 0 {\rm \ \ s.t.\ \ \ }
\g(u > u_1 -\e) \subset \CI^+(p_0) }
Thus, to prove \claimDpp, we need to find two separate relations: 
one delineating an appropriate region $R_{\d} \subset \CI^+(p_0)$
(which we will show must be entered by all causal curves $\g(u \to u_1)$),
and the other specifying the coordinate relations satisfied by
{\it any} causal curve $\g$ which asymptotes to the $u= u_1$ plane.
These relations can be written in terms of coordinate inequalities,
essentially relating $r$ and $v$.
Typically, they will take the form 
\eqn\genineq{
v \ge  \a \, g(r)}
 where $\a$ is 
an arbitrary constant and $g(r)$ is a particular function of $r$,
depending on the precise form of the metric.
For pp-waves with $F$ growing sufficiently slowly with $r$,
$g$ takes the form $g(r)= r^2$; generically $g(r) \sim F(r)$. 
If the inequality specifying $\g(u \to u_1)$ is more stringent than,
\ie\ implies, the inequality specifying $R_{\d}$ (or if they are identical),
then obviously any $\g(u \to u_1)$ must enter $R_{\d} \subset \CI^+(p_0)$.
This is pictorially sketched in Fig.4,  which denotes the $u-v$ plane.  
The dashed lines represent
the $u= u_0$ and $u=u_1$ planes; $p_0$ is any point on the former, 
while $\g$ represents any causal curve asymptoting to the latter.
We want to show that any such curve must eventually enter the region
$R_{\d} \subset \CI^+(p_0)$.

\ifig\figpp{Idea of proof of \claimBpp; specifically, any causal curve
$\g(u \to u_1)$ which asymptotes to the $u= u_1$ plane must enter into the
region $R_{\d} \subset \CI^+(p_0)$ in the causal future 
of any point $p_0$ with its $u$ coordinate $u_0=u_1 - \d$, for
arbitrarily small $\d$. 
The $u-v$ plane is shown on the sketch, but note that this is not a Penrose
diagram of a general pp-wave.}
{\epsfxsize=6cm \epsfysize=6cm \epsfbox{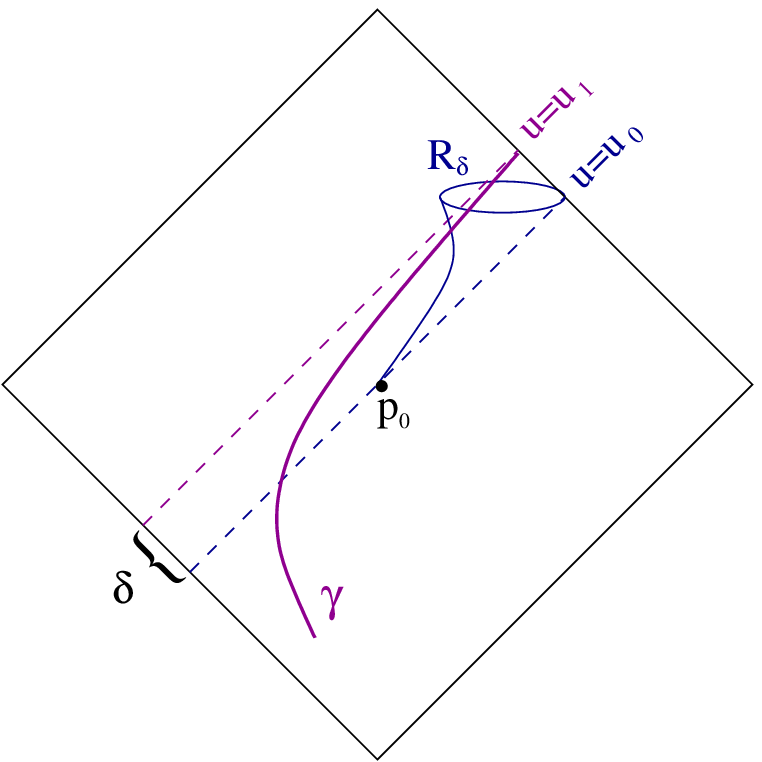}}

Let us first consider the relation on the coordinates far along any 
causal curve $\g$ which asymptotes to the $u=u_1$ plane.
As explained above, since $\g$ is future-directed and causal, 
$u$ must increase along it, so we can parameterize $\g$ by $u$.
Also, since we are interested in the part of the curve above the 
$u= u_0 = u_1 - \d$ plane, and $\d$ will eventually be taken arbitrarily
small, we can neglect the $u$-dependence in the metric, \ie, we can
take $F(u, r, \Omega) = F(u_0, r, \Omega)$ in \pparad.
From \pparad, the causal relation can then be written as
\eqn\causalgamma{
2 \, \dot{v} \ge \, \dot{r}^2 - F(u_0, r, \Omega) + r^2 \dot{\Omega}^2
}
where $\dot{} \equiv {d \over du}$.
Now, since $\g$ asymptotes to a finite-$u$ plane, some other coordinate
must diverge.  As for plane waves, it can be easily shown that 
$v(u) \to \infty$ as $u \to u_1$.  
This implies that $\vd \to \infty$ as $u \to u_1$ as well.
On the other hand, $r(u)$ can behave in several distinct ways; 
so we will now analyze these cases in turn.

\noindent
{\bf Case 1.}
$r(u)$ remains finite as $u \to u_1$.
Since $v$ diverges, this means that we can satisfy the inequality 
\genineq\ with $g(r) = r^p$
for any $\a$ and $p$, far enough along $\g$.

Alternately, in {\bf Case 2},
$r(u) \to \infty$ as $u \to u_1$.
Now it will be relevant how the function $F(u_0, r, \Om)$ behaves
as $r$ gets large.  There are, in turn, several distinct possibilities.

\noindent
{\bf Case 2a.}
$ \dot{r}^2 - F + r^2 \, \dot{\Omega}^2 
\longrightarrow  \dot{r}^2 + r^2 \, \dot{\Omega}^2 $ as $u \to u_1$,
\ie, as $r \to \infty$.
Then the causal relation \causalgamma\ may be reproduced by the 
 flat fiducial metric 
$ds^2 = -2 \, du \, dv +\( 1 - \bar{\e} \) \(dr^2 + r^2 \, d\Om^2\)$.
This can be integrated to yield the finite difference relation
along $\g$,
\eqn\findiffa{
2 \, \Delta v \, \Delta u \ge \[ (\Delta r)^2 + r^2 \, (\Delta \Om)^2 \]
\, (1-\tilde{\e})}
where $\tilde{\e}$ 
encodes the error in neglecting $F$, which can be made arbitrarily 
small by going far enough along $\g$.
Now, for large $v$, $ \Delta v \approx v$, which again becomes arbitrarily
accurate by going far enough along $\g$.
Also, since this inequality is most stringent if the final $r$, and therefore
$\Delta r$, is large, we can also take $ \Delta r \approx r$.
Thus \findiffa\ yields the inequality
\eqn\appfda{
v \ge {(1 - \tilde{\e}) \over 2 \, \Delta u} \[1 + (\Delta \Om)^2\]  \, r^2}
Since $(\Delta \Om)^2$ is a finite ${\cal{O}}(1)$ number, we can 
rewrite this as 
\eqn\ineqa{
v \ge \a \, r^2}
where $\a$ can be arbitrarily large by choosing $\Delta u$ sufficiently
small.  In the asymptotic region along $\g$, this is directly related
to choosing $\e$ sufficiently small in \claimDpp.

\noindent
{\bf Case 2b.}
$ \dot{r}^2 - F + r^2 \, \dot{\Omega}^2 
\longrightarrow   - F > 0 $ as $u \to u_1$.
The inequality on $F$ should be taken to mean the inequality for
the asymptotic region along $\g$, \ie, $F(u_0, r \to \infty, \Om_{\g})$,
which has a definite sign.  (The sign of $F$ may change 
along different curves $\g$ which reach infinite $r$ in different directions,
\ie\ for different $\Om$.)
Then \causalgamma\ implies that
$2 \, \vd \ge -F \gg \dot{r}^2 + r^2 \, \dot{\Omega}^2$, 
and the same bound \ineqa\ as derived above must apply.
However, this is not sufficient; we can in fact derive a much
stronger bound.  
By considering the corresponding finite difference relation, 
$2 {\Delta v \over \Delta u} \ge |F|$,
we arrive at the relation
\eqn\ineqb{
v \ge \e' \, |F(u_0, r, \Om)|} 
where $\e'$ is small (corresponding to small $\Delta u$), but fixed.

\noindent
{\bf Case 2c.}
$ \dot{r}^2 - F + r^2 \, \dot{\Omega}^2 
\longrightarrow   - F < 0 $ as $u \to u_1$.
When $F>0$, we encounter the most trivial case.
This is because in the relevant regions, causality in our spacetime 
\pparad\ is more stringent than in the flat spacetime.
So by the same type of argument as in \nobh, we can immediately
see that the claim
${\rm TIP}\[\g(u \to u_1)\] = \{(u,v,r,\Om): u \le u_1 \}$ 
must be satisfied.

\noindent
Note that when ${ \(\dot{r}^2  + r^2 \dot{\Om}^2 \) \over |F|} \to {\rm 
const} $ as $ r \to \infty$, as in the plane wave case, we can use the 
same construction as in Case 2a or Case 2b, with appropriate modifications 
of the constants involved.
 
To summarize, we have obtained two relations between the coordinates
far enough along any causal curve which asymptotes to the $u = u_1$ plane,
given by \ineqa\ and \ineqb\ in the respective cases depending on the 
form of $F$ in the pp-wave spacetime \pparad.

We now turn to the second part of our proof, namely constructing
the region $R_{\d}$ in the causal future of a point $p_0$ lying on the
$u = u_0$ plane.
Since we want to ascertain that all the curves considered above enter
this region, it suffices to consider the part of the causal future of 
$p_0$ with $u$ between $u_0$ and $u_1$, and the coordinate $v$ large.
To that end, we may again make the approximation $u=u_0$ in the 
metric \pparad.

We want to show that if at some point $p_f = (u_f,v_f,r_f,\Om_f)$,
the inequality \ineqa\ or \ineqb\ is satisfied (for $u_f$ sufficiently 
close to $u_1$ and $v_f$ sufficiently large),  then $p_f \in \CI^+(p_0)$.
To this end, it suffices to construct a causal curve $\C$
from $p_0$ to $p_f$.
We will construct such a curve in three stages, $\C = \{ \C_1,\C_2,\C_3 \}$,
by introducing two
convenient intermediate points, $q_0$ and $q_f$.  
In particular, $q_0$ will have the final angle, $\Om_f$, but $r=r_0$,
while $q_f$ will be chosen
so as to saturate the inequality \ineqa\ or \ineqb, and all coordinates
except for $v$ will match those of $p_f$.
Pictorially, we construct causal curves between respective points
as follows:
\eqn\curve{\eqalign{
 p_0 &= (u_0,\ v_0,\ r_0,\ \Om_0) \cr
\downarrow \ & \C_1 \cr 
 q_0 &= (u_q,\ v_q,\ r_0,\ \Om_f) 
{\rm \ \ \ with \ } u_0 < u_q < u_f ,
{\rm  \ and \ } v_0 < v_q < v_f \cr
\downarrow \ & \C_2 \cr 
 q_f &= (u_f,\ v=\a \, r_f^2,\ r_f,\ \Om_f) 
\ \ \ {\rm for \ case \ 2a,}\cr 
& \ \ \ {\rm (or \ \ } q_f = 
(u_f,\ v=\e' \, |F(u_0, r, \Om_f)| ,\ r_f,\ \Om_f)
 \ \ \ {\rm for \ case \ 2b)}\cr 
\downarrow \ & \C_3 \cr 
 p_f &= (u_f,\ v_f,\ r_f,\ \Om_f)
}}
where the constants 
 appearing in $q_f$ are to be picked later.
It is easiest to construct the last curve, $\C_3$, from $q_f$ to $p_f$,
because  $v$ is the only
coordinate along $\C_3$ which needs to change. Since $\dva$ is
a null Killing field, it is readily apparent that $q_f$ and $p_f$ are 
in fact connected by a null geodesic following the orbit of $\dva$.

Let us now proceed with constructing  $\C_1$, the curve from $p_0$ to $q_0$,
which takes us around the origin.
To this end, we can choose the coordinates along $\C_1$ to satisfy
\eqn\curveOm{\eqalign{
\rd & = 0
  \ \ \ \Rightarrow \ \ \ F=F(u_0,r_0,\Om) \equiv F_0(\Om) \cr
2 \, \vd & = A  \cr
r_0^2 \, \dot{\Om}^2 & = A + F_0(\Om)
}}
where $A$ is some positive constant, chosen such that a solution
exists.  In particular, since the LHS of the last equation must be 
positive, whereas $F_0$ can be negative,
 we require that $A>- F_0(\Om)$.  This is possible, since 
$F_0(\Om)$ remains bounded at any constant $r$.
Clearly, such a curve $\C_1$ satisfying \curveOm\ exists,
and from  \causalgamma, it is null and therefore causal.
The remaining requirement on $\C_1$ is that $u_q$ lie between $u_0$ and
$u_f$, \ie, that $\C_1$ rounds the origin sufficiently fast; but this is 
easily satisfied.  In fact, we can make $u_f$ arbitrarily close to 
$u_0$ by taking $A$ large enough.

Finally, we need to construct the intermediate curve $\C_2$ between $q_0$ 
and $q_f$, which takes us to large $r$ while saturating the requisite
inequality on $v$.
This can be achieved by the following conditions:
\eqn\curver{\eqalign{
\dot{\Om} & = 0 \ \ \  \Rightarrow \ \ \  \Om = \Om_f \cr
B \, \dot{r}^2 &= F(u_0,r, \Omega_f) \equiv f(r) \cr
2 \, \dot{v}  & = (1 - B) \, \dot{r}^2 = { (1-B) \over B} \, f(r)
}}
where $B$ is some constant, $B<1$.  
Note that, unless $f=0$, by taking $B$ sufficiently small, our
curve $\C_2$ can reach\foot{
Construction of such a curve could be problematic in 
regions where $f(r)$ changes sign. To specify the curve completely,
we  therefore pick the sign of $B$ depending on the sign of $f(r)$
near $r = r_0$, such that $\dot{r}^2 \ge 0$. 
Now suppose $f(r)$ changes sign for some $r=r_i \, > \, r_0$.
At $r_i$ we also flip the sign of $B$ and continue with the construction 
of the causal curve. In other words, we can always solve the equation
$\dot{r}^2 = |f(r)/B|$, for which the existence of a solution is
guaranteed. Of course, such a curve may
 have higher derivative discontinuities; but we can 
then smooth these out.} 
sufficiently large values of $r$ and correspondingly $v$.
Now we can repeat the same finite-difference argument as above:
for instance, in case 2a,
for sufficiently small $\Delta u$, the last equation in \curver\
implies that 
$2 \, \Delta v \, \Delta u = (1-B) \, (\Delta r)^2 \, (1-\tilde{\e})$,
so that choosing $\a = {(1-B)  \, (1-\tilde{\e}) \over 2 \, \Delta u}$,
we obtain the necessary relation \ineqa\ for $v$ in terms of $r$.

We can see the above construction more explicitly by considering several 
illustrative examples.
Let us first confirm the validity of $\C_2$ for a simple plane wave
example, $F(u_0,r,\Om_f) = r^2$.  Then \curver\ implies that
$r(u) = e^{u/\sqrt{B}}$ and 
$v={(1-B) \over 4 \sqrt{B}} \,  e^{2u/\sqrt{B}} = \a \, r^2$ for
$\a \equiv {(1-B) \over 4 \sqrt{B}}$, which is
clearly consistent with \ineqa.
As a second example, let us consider the more ``dangerous'' type 
of behavior, such as $F(u_0,r,\Om_f)= - r^4$.  To simplify notation,
let us pick the arbitrary constants $r_0= 1$ and $u_0 = 0$.
Then $r(u) = {\sqrt{|B|} \over \sqrt{|B|} - u}$, 
so that from integrating the
last equation of \curver\ and re-expressing $v(u)$ in terms of $r$, 
$v= {|B| \, (1-B) \over 6} \, r^3$.  
Clearly, for sufficiently small $|B|$, we can satisfy 
${|B| \, (1-B) \over 6} \le \e' \, r_f$, which yields a relation 
consistent with the necessary inequality \ineqb. 
As a final example (also in the 2b category), when $F \sim  - e^r$,
\curver\ gives $ v = \b(B) \, e^{r/2} \ll \e' \, e^r$ for 
$\b(B) \ll \e' \, e^{r_f/2}$. Again, this implies that any causal curve
which asymptotes to the $u=u_1$ plane, that necessarily satisfies 
$v \ge \e' \, e^r$, enters into the future of $p_0$, given by 
$v \ge \b(B) \, e^{r/2}$. 

Thus, we have seen that by putting together\foot{
To obtain a smooth curve, we would need to smooth out the sharp edges.
Typically, this can be done in such a way that the new curve is 
timelike, since we can always perturb the curves to introduce a small
timelike component.}
three null curves, $\C_1$, $\C_2$, and $\C_3$,   we can construct
 a causal curve $\C$ which connects $p_0$ with $p_f$.
The points $p_f \in R_{\d}$ by construction
all satisfy the requisite relation
insuring that all causal curves which asymptote to $u=u_1$ must 
enter $R_{\d}$.  
This completes the proof of the claim \claimApp, that 
${\rm TIP}\[\g(u \to u_1)\] = \{(u,v,r,\Om): u \le u_1 \}$.
{\bf (QED)}

\subsec{Identifications for general pp-waves}

The preceding subsection established the claim \claimApp\ that the
TIP of any causal curve which asymptotes to finite $u_1$ is given by the
set of all points with $u \le u_1$.
By time-reversing our statements, the analogous result holds for the 
TIFs.
Thus, this part of $\scri$ is parameterized by a single parameter $u$,
and is therefore one-dimensional.
However, as cautioned in Section 4.4, this does not automatically
imply that the full causal boundary of pp-waves is one-dimensional,
as demonstrated by the Minkowski spacetime example. Note 
however, that the construction of TIPs for general pp-waves at
finite $u$ enables us, as in Sec 4.2, to infer the absence of 
horizons in these spacetimes, reconfirming the claims of \nobh. 
What remains to be examined, in order to determine the full causal
structure, are the identifications between the TIPs and TIFs.

In order to ascertain which ideal points need to be identified, based on 
causal properties alone, we need to isolate the ideal points which have the  
identical causal past and future. Let us start by considering ideal points 
which correspond to TIPs. Here, since the ideal points are labelled by 
a single coordinate $u$, it is clear that there are no identifications between 
the ideal points. Considering two values of $u$, say $u_1$ and $u_2$,
with $u_1 < u_2$ it is 
clear that the causal past of all causal curves asymptoting to $ u = u_1$ 
is distinct from that of causal past of curves asymptoting to $ u= u_2$.
By time-reversal, similarly there are no identifications between the ideal 
points corresponding to TIFs. 

The main issue then is whether there exist any identifications
between the ideal points associated with TIPs and TIFs.
 To this end, we can 
employ the same technique as for the plane wave case. To wit, if there exists 
a sequence of causal curves which emanate from {\it any} point in the 
$ u = u_0$ plane, say $p_0 = (u_0,v_0,x^i_0)$, and have an 
accumulation curve which reaches the point $p = ( u _1, v = -\infty, x^i)$, 
where $x^i$ are arbitrary but finite, and
$u_1 > u_0$ is the smallest value of $u$ for which such an accumulation curve
exists, then we would identify 
${\rm TIP}\[\g(u \to u_0)\]$ with the  ${\rm TIF}\[\g(u \to u_1)\]$. If 
such identifications persist for arbitrarily large values of $u$, then we see
that the structure at $u = \infty$ is different from \eg\ that of
 Minkowski space, 
and we will have a one-dimensional causal boundary. If we 
encounter singular behaviour of the functions $F(u,x^i)$ at finite values of 
$u = u_\infty$, then the ideal points associated with the $u = u_\infty$ plane 
would correspond to a null singularity. Also there are situations with 
regular behaviour of $F(u,x^i)$ with respect to $u$, but without 
any identifications.

To complete the program of the causal structure for pp-waves, we therefore 
need to understand in which cases are there appropriate causal curves which 
allow for identifications between the ideal point associated with a TIP 
and a TIF.  To this end, it is useful to analyse in detail the behaviour 
of the geodesics in the pp-wave spacetimes. If we can  establish that the 
behaviour of the geodesics is oscillatory for arbitrarily large values of $u$, 
then, as in the plane wave case, with a clever choice of 
initial conditions we can 
establish that causal communication to arbitrarily 
large negative values of $v$ is possible. Note that in the absence of 
caustics, null geodesics bound the future light cone emanating from a point.
Hence, if we can show that it is not possible for null geodesics to 
reach large negative $v$, then causal curves will also be unable to 
reach $ v \to -\infty$.

The derivation of the geodesic equations is presented in the 
Appendix A; we have 
\eqn\geodxpp{
\xdd^i + {1 \over 2} \p_i F(u, x^i)  = 0}
\eqn\geodvpp{
2 \vd = - F(u, x) + \sum_i( \xid)^2 }
Note that unless $F(u, x)$ takes the plane wave form $f_{ij}(u) \, x^i x^j$,
$\vd$ is not a total derivative, so we can't write a general
formula for $v$ analogous to \veq, without first solving the 
$x$-equations \geodxpp. 
In what follows, we will analyze these geodesic equations for certain special 
cases and show that, while there are identifications in certain examples,
such as the pp-wave background discussed in \malmaoz\  with metric as 
given in \mm, generic
vacuum 
pp-waves do not admit any identifications and are singular
(geodesically incomplete). We will first consider examples where 
we have identifications between the ideal points, as they are most 
analogous to the hitherto analyzed plane wave examples.

\vskip3mm
\noindent
{\bf PP1.} Our first example is the pp-wave background \mm,
where $F(u,x^i) \equiv \cosh x -  \cos y$. This is 
interesting from the string theory point of view since it leads to an 
integrable sine-Gordon 
like theory on the world-sheet in light-cone gauge. The geodesic equations
in this 
case need to be analysed only for the coordinates, $x$ and $y$, 
$F(u,z^i)$ is independent of the remaining coordinates.
The geodesic equations 
\geodxpp\ read, 
\eqn\mmgeod{\eqalign{
\ddot{x} + {1 \over 2} \, \sinh x & = 0 \;\;\; \Rightarrow \;\;\;  
\dot{x}^2 + \cosh x = \a \cr  
\ddot{y} + {1 \over 2} \, \sin y & = 0 \;\;\; \Rightarrow \;\;\; 
\dot{y}^2 - \cos y = \b ,
}}
with arbitrary constants of integration $\a$ and $\b$. Given the 
solution to \mmgeod, we can determine $v(u)$ from the equation
\eqn\mmgveq{
2\, \vd = - F(x,y) + \dot{x}^2 + \dot{y}^2 = \a + \b - 2 \cosh x(u) + 
2 \cos y(u). }

Clearly, the geodesics in the $y$ direction are not going to be oscillatory 
 for $\b > 1$, 
while those in the $x$ direction are going to oscillate. The 
geodesic motion in the $x$ direction is just the motion of a particle in a 
$\cosh x$ potential, with a fixed total energy $\a$. Given a fixed energy, the
particle can rise up to the potential upto the level specified by the energy 
and then starts to roll back down. Since the potential is reflection symmetric,
this process continues indefinitely. 

\ifig\Figmm{Plot of $v(u)$ for \mm\ spacetime for 
null geodesics moving in $v$ and $x$ directions only. $\a$ in \mmgeod\
is chosen to be $\cosh 3$ and $\cosh 5$. The $\a = \cosh 5$  
curve attains larger values of $|v(u)|$. }
{\epsfxsize=9cm \epsfysize=6cm \epsfbox{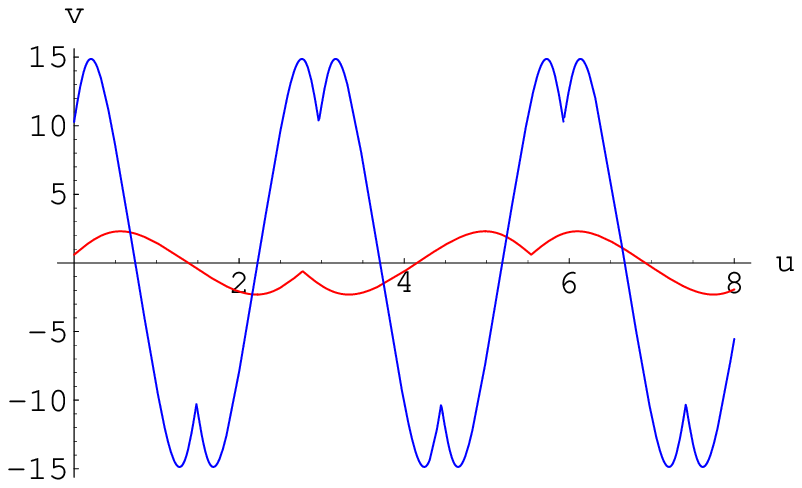}}

From any point $p_0 = (u_0,v_0,x_0,y_0,z^i_0)$ in the spacetime \mm,
we can find a sequence of null geodesics $\g_n$, parameterised by 
$\a_n$, such that along $\g_n$ we have $v_n \to -\infty$. This is 
made possible by taking $\a$ arbitrarily large in the geodesic equations
\mmgeod, \mmgveq. We illustrate this in Fig.5, where we plot 
$v$ as a function of $u$, for two different values of $\a$. The higher
curve corresponds to the larger value of $\a$.
Thus, it is possible in this example 
to find identifications between the TIPs and the TIFs and in particular, we 
can claim that
 {\it the causal boundary of the pp-wave \mm\ is one-dimensional.}

As for the BMN plane wave case, one can also show that this one-dimensional
boundary is locally null.  This can be achieved by demonstrating that 
two ideal points $P$ and $Q$ are causal, \ie, there exists a sequence of
points $P_n \to P$ with all $P_n$ timelike separated from $Q$, but
there exists a sequence of points $Q_n \to Q$  with all $Q_n$
spacelike separated from $P$.

Before proceeding to other examples, we wish to point out an interesting 
feature associated with this example. It is a common belief that plane 
waves can be the only geodesically complete pp-waves. This relates to 
an unproven conjecture of Ehlers and Kundt \ek\  
as stated in \bicak. The pp-wave \mm\ is, however, 
{\it geodesically complete}.\foot{
In fact, we will see later that there do exist vacuum pp-waves which
are geodesically complete in $d \ge 5$ dimensions.}
From the geodesic equations \mmgeod\ it is clear that the confining nature 
of the potential in the $x$ direction keeps the geodesic from running off to 
infinity at finite affine parameter. Along the $y$ direction, one has 
oscillations 
with finite amplitude superposed on linear growth with time, which also 
doesn't asymptote to infinity at finite values of the affine parameter.

The solutions to the geodesic equations \mmgeod, may be written down 
explicitly in terms of the Elliptic functions as:
\eqn\mmgeodesics{\eqalign{
z^i(u) &= \kappa \, (u + u_0) \cr 
u + u_0 & = -{2 i \over \sqrt{\a -1} } \, 
{\rm F}\( { i \, x(u) \over 2} , -{2 \over \a -1}
\) \cr
u + u_0 & = {2 \over \sqrt{1 + \b} } \, 
{\rm F}\( { y(u) \over 2} , \, {2 \over \b +1}
\) }}

\eqn\mmgevsol{\eqalign{
v(u) &= \( \kappa^2 + \eta \) ( u+ u_0) +  \cr 
& \hskip5mm -{2 i \over \sqrt{\a -1} } \, 
\[ \(\a -1\) \, 
{\rm E}\( { i \, x(u) \over 2} , -{2 \over \a -
1} \) - { \a \over 2} \, {\rm F}\( { i \, x(u) \over 2} , -{2 \over \a -1}
\) \] \cr 
& \hskip8mm + {2 \over \sqrt{1 + \b}}  \, \[  \(\b +1 \) \, 
{\rm E} \( { y(u) \over 2} , {2 \over \b +
1} \) - { \b \over 2} \, {\rm F} \( { y(u) \over 2} , {2 \over \b +1}\) \].
}}
In the above $\eta = (\pm 1,0)$ denote timelike, spacelike, and null 
geodesics, respectively, and ${\rm F}(\phi,m)$ and $
{\rm E}(\phi,m)$ are the elliptic integrals of 
the first and second kind, respectively \gradr.

\vskip3mm
\noindent
{\bf PP2.} Motivated by the example in {\bf PP1}, we can try to
characterize the cases with 
identifications. Let us assume for simplicity that $F(u,x^i)$ 
is independent of $u$ and can be written as 
sum of functions, each of which depends on  
single coordinate \ie, $F(u,x^i) = \sum_j \, f_j(x^j)$. 
Essentially we want $\p_i \p_j F(u,x^i) =0$ and $\p_u F(u,x^i) =0$. 
A prototypical example of such a spacetime is of course \mm.
The idea is to decouple the geodesic equations \geodxpp\ so that 
they may be analyzed separately. If we can find 
oscillatory behaviour along any one of the coordinates, then  
we will be able to conclude that there are identifications between 
TIPs and TIFs, unless we are hampered by some of the other 
coordinates leading to singularities. 

We concentrate on a single coordinate, 
$x \in (-\infty, \infty)$, and write the corresponding function as $f(x)$. 
We also assume that all other coordinates are sufficiently 
well behaved \ie, the geodesics in these directions are complete. 
Concentrating on geodesics along $x$, 
geodesic equations \geodxpp\ and \geodvpp\ read
\eqn\splppgd{\eqalign{
\ddot{x} + {1\over 2} \, {\partial f(x) \over \partial x }  = 0 & \;\;\; 
\Rightarrow 
 \;\;\; 
\dot{x}^2 + f(x) = \a \cr
 2\, \dot{v}   = \dot{x}^2 - f(x) & \;\;\; \Rightarrow \;\;\; 
\dot{v} = {\a \over 2} - f(x) 
}}
with $\dot{}  \equiv {d \over du}$. 
We can interpret the geodesic equations as motion of a particle along the $x$ 
direction in a potential $f(x)$ and total energy $\a$. 

From the first equation in \splppgd\ it is clear that as long as $f(x)$ is 
bounded from below, the particle doesn't reach $x = \infty$ in finite time 
$u$. Therefore such spacetimes are 
geodesically complete.  This, however, does not mean that 
spacetimes with $f(x)$ not bounded from below are  
geodesically incomplete. For example, when 
$f(x) = -x^2$, the particle trajectory as a function of time is exponential, 
so it takes infinite time to reach out to the asymptotic regions. 
If $f(x)$ is bounded from below and in addition  
$f(x \to \pm \infty) \to + \infty$, then there are 
identifications between the 
TIPs and TIFs, and the causal boundary is one-dimensional. 

To exemplify the situation, consider the case when $f(x) = x^{2n}$ for 
some integer $n >1$.
In this case we have the geodesic equation $ \dot{x}^2 + x^{2n} = \a$, 
which has oscillatory 
solutions. In particular, $u = { x(u)  \over \sqrt{\a} } \, 
{\rm F}\({1\over 2n} 
, {1\over 2 } ;
1+ {1 \over 2n}; {x(u)^{2n} \over \a} \)$ with ${\rm F}(a,b;c;x)$ being the 
hypergeometric function and $v(u)$ can be expressed implicitly
as a function of $u$. One can verify 
numerically that the oscillatory 
behaviour implies identifications between the TIPs and TIFs for 
arbitrarily large values of 
$u$, leading to a one-dimensional causal boundary.

On the other hand, if one were to choose $f(x) = x^{2n +1}$ for some integer 
$n> 1$, then the 
spacetime is geodesically incomplete. This is because there is an 
instability for $x < 0$ and the 
potential $f(x)$ is unbounded from below. So the fiducial particle moving in 
this potential is 
going to run away to infinity in finite time. 
However, since along these geodesics $ v \to + \infty$, 
there aren't any causal curves or geodesics in these spacetimes 
which will 
allow us to causally communicate from any point $p_0 = (u_0, v_0, x_0)$ to 
$p = (u_1, v = -\infty, x)$ with finite $x$ and $u_1 > u_0$. 
From this we conclude that the ideal points correspond to 
singularities and that there are no identifications between them.

\vskip3mm
\noindent
{\bf PP3.} Our next example will be vacuum pp-waves, which we will take to be 
of the form presented in \pparad\ and \F. We will show that vacuum pp-waves 
are generically geodesically incomplete; the exceptions being the flat space 
and plane waves \plane\ with regular functions $f_{ij}(u)$,
and the monopole pp-wave in dimension $d \ge 5$.
As we see from \F, one can have singular behaviour of $g_{uu}$ in two 
distinct ways; either the 
metric component is ill-behaved as $r \rightarrow \infty$, or 
there is a singularity as $r \rightarrow 0$. Since we can superpose 
solutions, in generic vacuum pp-waves both such behaviours are 
possible. We will investigate two cases, wherein the 
singularities are either only at $r=0$ or only at $ r =\infty$.

Consider vacuum pp-waves \pparad\ with the function $F(u,r,\Om)$ as in \F, 
with $f_L^+(u) \equiv 0$ and $ f_L^-(u) \neq 0$ for some $L = \l_1 \ne 0$. 
For simplicity we further assume that ${d \over du} f^-_{\l_1}(u) = 0$. 
A particular example of such a spacetime would be
\eqn\fdppzero{
ds^2 = -2 \, du \, dv - {1 \over r^4} \, \(5 \cos^3 \theta - 3 \cos \theta \) 
 \, du^2 + dr^2 + r^2\, \(d\theta^2 + \sin^2 \theta d \phi^2 \) }
In \fdppzero\ there are geodesics which hit the singularity $r=0$ at a 
finite affine 
parameter, causing the spacetime to be geodesically incomplete. Furthermore, 
since the spacetime rapidly approaches flat space as $r \to \infty$, we find 
that there is no oscillatory behaviour of null geodesics, 
implying that there exists no reason from 
causal properties to identify ideal points corresponding to different TIPs 
with those 
corresponding to TIFs. This is the generic behaviour of vacuum pp-waves in 
arbitrary dimension so long as $f^+_L(u) \equiv 0$, with one notable
exception.  The monopole solution with $F(u,r,\Om) = +{1 \over r^{d-4}}$
in $d$-dimensions is geodesically complete. This must be so since
$F(r)$ is bounded from below.  One can check explicitly that geodesics
do not end at a finite affine parameter.  On the contrary taking
$F(u,r,\Om) = -{1 \over r^{d-4}}$ leads to geodesically incomplete
spacetimes.
In Appendix B we present some solutions to the geodesic equations \fdppzero. 

Now let us turn to the example when $f_L^-(u) \equiv 0$ and 
$ f_L^+(u) \neq 0$. We will 
in addition also assume ${d \over du} f^+_{L}(u) = 0$. As an example, one can 
take the five-dimensional vacuum pp-wave
\eqn\fdppone{
ds^2 = -2 \, du \, dv - r^3 \, \(5 \cos^3 \theta - 3 \cos \theta \) 
 \, du^2 + dr^2 + r^2\, \(d\theta^2 + \sin^2 \theta d \phi^2 \) }
If we try to look for geodesics at a constant value of the angular 
coordinate, \ie, geodesics with only a radial component, we find that they 
exist for $ \theta = 0$ or for $ \cos \theta = \pm \sqrt{ {1 \over 5}}$. 
Consider the radial null geodesic 
from $(u=0, v_0, r_0, \theta_0, \phi_0)$, with constant values of 
the angular variables given by arbitrary $\phi_0$
and $ \cos \theta_0 = \sqrt{ {1 \over 5}}$. This radial null geodesic 
reaches infinite values of the radial coordinate in finite affine parameter. 
On the other hand, a radial null geodesic sitting at constant angular 
variables $(\theta_0 =0, \phi_0 )$ with arbitrary $\phi_0$ exhibits 
oscillatory behaviour.
Such is immediately obvious by just considering the function 
$F(r,\theta) =  r^3 \, \(5 \cos^3 \theta - 3 \cos \theta \)$ at 
constant values of $\theta = \theta_0$. It is clear that for all values of 
$\theta_0$ such that $\cos^2 \theta_0  < {3 \over 5}$, the  radial potential
$F(r, \theta_0)$ is essentially negative definite. First of all, the 
presence of at least one geodesic which diverges at a finite affine 
parameter implies that the spacetime is geodesically incomplete. Furthermore, 
in order for us to claim the existence of identifications between the ideal 
points, it is not sufficient to find special null geodesics which 
exhibit oscillatory behaviour. What we need to show is that starting 
from {\it any} point on a plane of constant $u=u_0$, we can causally 
connect to a plane of constant $u=u_1$ with $u_1 > u_0$ with the coordintate 
$v \to - \infty$. 
This is not possible for generic geodesics in the 
spacetime \fdppone. 
Hence, we conclude that there are no identifications. 

\vskip3mm
\noindent
{\bf PP4.} Our last example is the four dimensional vacuum pp-wave spacetime
\foot{We thank Chris Hillman for this example.}
\eqn\hill{
ds^2 = -2\, du \, dv - \sin x \, e^y  \, du^2 + dx^2 + dy^2
}
The geodesic equations \geodx\ in this spacetime read
\eqn\ghill{\eqalign{
\ddot{x} + {1 \over 2} \cos x \, e^y & =0 \cr
\ddot{y} + {1\over 2} \sin x \, e^y & = 0
}}
Because the function 
$F(x,y) = \sin x \, e^y$ is not a positive definite function in the 
variable $x$, we find 
that the potential for motion in the $y$ direction is 
unbounded from below. This causes the geodesics to diverge at 
finite values of $u$. 
In Fig.6 we plot the geodesics $(x(u),y(u))$ for some generic 
initial conditions to illustrate the point.
 
\ifig\Fighill{Geodesics $\{x(u),y(u)\}$ for the spacetime \hill. We 
present solutions to \ghill, for some generic initial conditions to 
exhibit geodesic incompleteness.
}
{\epsfxsize=8cm \epsfysize=6cm \epsfbox{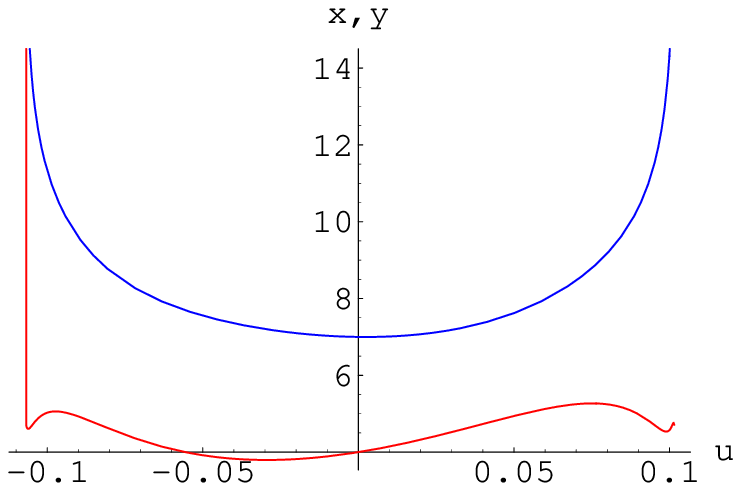}}

\vskip3mm
\noindent
From examples {\bf PP3} and {\bf PP4}, it is clear that for vacuum 
pp-waves there 
are no identifications between TIPs and TIFs. The reason that this happens is 
that for vacuum solutions we need the function $F(u,x^i)$ in \pp\ to 
be harmonic. The only harmonic function that is bounded from below
is the constant function (or the monopole solution),
\ie, with no angular dependence. All other harmonic 
functions have domains where they either diverge to $+ \infty$ or to 
$- \infty$. The divergence towards 
positive infinity is acceptable, since this leads to confining potentials;
but the divergence towards negative infinity causes runaway behaviour 
and thence generically 
to geodesic incompleteness. The situation is somewhat mitigated if 
$F(u,x^i)$ grows slower than $(x^i)^3$ 
in the transverse coordinates (here we are 
assuming that  $F(u,x^i)$ is a regular function of $u$; if it is not we have 
null singularities at $u = u_\infty$, where $F(u \to u_\infty,x^i) \to 
\pm \infty$). For $F(u,x^i) = a(u) + b_i(u) \, x^i$  
we have flat space, and for 
$F(u,x^i) = f_{ij}(u) \, x^i \, x^j$  we have plane waves,  
which are geodesically complete. 
Geodesic completeness for generic plane fronted waves with 
sub-quadratic growth of $F(u,x^i)$ was recently discussed in \candela.

\newsec{Discussion}

In the present paper we have analyzed in some detail the 
causal structure of pp-wave spacetimes. This work generalises 
the recent discussion of causal structure of certain plane wave spacetimes 
\bena,\maro. Since most pp-wave spacetimes 
are not conformally flat, one cannot use the simple trick of 
reading off the causal structure from the conformal
rescaling into Einstein Static Universe.
Instead, as \maro, we use the technique developed by 
Geroch, Kronheimer, and Penrose \geroch, which identifies the causal 
structure by looking at future(past) endless causal curves in the 
spacetime and assigns ideal points to these curves. The ideal 
points have the same causal future(past) as the curve. Having 
assigned ideal points to the spacetime manifold, one has to ensure that 
this assignment is minimal. This requires that different ideal points
have distinct causal domains of influence. 
A more stringent requirement is that upon adjoining the 
minimal set of ideal points to our spacetime, we obtain a smooth 
Hausdorff manifold.

As we have seen, for plane waves where the metric component 
$g_{uu}$ doesn't approach flat space too rapidly, 
such as polynomials, trigonometric, or hyperbolic 
functions $f_{ij}(u)$ in \plane, the causal boundary 
is one-dimensional and null. On the other hand, spacetimes where 
the approach to flat space is sufficiently rapid in $u$, the 
causal boundary is higher dimensional, as in Minkowski space. 
The plane wave spacetimes 
obtained as Penrose limits of supergravity backgrounds dual to 
non-local theories, such as little string theory and non-commutative 
gauge theory, generically have a one-dimensional causal boundary which 
is null. 
In the case of non-commutative theories, in certain corners of  
parameter space, the spacetime becomes singular and loses the 
property of having a one-dimensional boundary.  

For pp-wave spacetimes \pp, the situation is more diverse, since now 
the function $F(u,x^i)$ can exhibit singular behaviour in the transverse 
coordinates as well as in $u$. 
For the spacetime \mm\ discussed in 
\malmaoz, we find that the causal boundary is one-dimensional and null.
More generally, for geodesically complete 
pp-wave spacetimes wherein the geodesic equations decouple, we have 
a one-dimensional causal boundary if the following is satisfied: 
$F(u,x^i) = \sum_j \, f_j (x^j)$ and at least one of the $f_j$'s diverges 
to plus infinity as $x^j \to \pm \infty$, in other words, oscillations 
of $x^j(u)$ induce an oscillatory behaviour in $v(u)$. 
In the case of generic 
vacuum pp-waves which are not plane waves or flat space, 
we find that the spacetimes are geodesically incomplete and 
the causal boundary is no longer one-dimensional. In addition to explicitly
constructing the causal structure for these spacetimes, we have demonstrated 
that there are no horizons, since any point in these spacetimes is 
causally connected to infinity, thereby putting the arguments of \nobh\ at a 
level of greater rigor.

One interesting aspect uncovered in our analysis relates to 
geodesic completeness of pp-waves. The common understanding is that 
plane waves are the only geodesically complete subset of pp-waves
\host,\bicak,\ek. We have 
shown that the monopole vacuum pp-wave and some
 non vacuum pp-waves, such as \mm, are geodesically complete. In fact, 
we can demonstrate geodesic completeness
for pp-wave spacetimes wherein the geodesic equations decouple,
and the functions appearing in $g_{uu}$ are bounded from below for 
each of the transverse space coordinates. This criterion is 
satisfied if, for 
$F(u,x^i) = \sum_j \, f_j (x^j)$, the functions 
$f_j(x^j)$ are either bounded 
from below or $|f_j(x_j)|$ grow slower than $(x^j)^3$. 
Interestingly, the geodesic equations \geodxpp\ are identical to the 
world-sheet equations of motion, for strings propagating in the 
background \pp. For a well defined world-sheet sigma model, 
we would require that the `potential function' $F(u,x^i)$ be bounded 
from below. The connection between classical equations of motion of the 
string world-sheet, and geodesic equations for null geodesics
in the spacetime, is quite suggestive of the fact 
that the string propagation in 
geodesically incomplete pp-wave backgrounds suffers from some pathological 
behaviour.

In general relativity, a geodesically incomplete spacetime,
\ie, a spacetime wherein at least one in-extendible geodesic ends at
a finite value of its affine parameter, is by definition a singular
spacetime. 
Conversely, if all geodesics can be shown to be complete, as in the
above example, such a spacetime is not singular, despite some 
Riemann curvature tensor components diverging.
In the cases when there are singularities, the usual expectation 
is that we cannot extend the 
spacetime past a singularity, since observers cannot pass through.
It is interesting to note that the last statement, however, need not be true.
As we have seen in the vacuum pp-wave spacetimes, 
there can be special geodesics with very non-generic properties.
As a simpler example, consider the (non-vacuum) plane wave spacetime
\plane\ with $f_{ij}(u) = {1 \over u^2} \, \d_{ij}$, \ie,
\eqn\singpl{
ds^2 = -2 \, du \, dv - {1 \over u^2} \, (x^i)^2 \, du^2 + dx^i \, dx^i.}
The corresponding geodesic equation is 
$\xdd^i + {1 \over u^2} \, x^i  = 0$, so that generic geodesics end at
$u=0$.  However, the special geodesic with initial conditions 
$x^i(u_0)= 0$ and $\xd^i(u_0)= 0$ can be extended through $u=0$.
In fact, we can use the planar symmetry of plane waves to generate a whole
family of such geodesics.
However, for a physical observer, the tidal forces still diverge at
$u=0$ in this example.

Singular plane wave geometries, such as discussed above,
may provide useful toy models for studying singularity resolution
in string theory.  
In particular, there are two useful features of these spacetimes.
Firstly, they can be viewed as a Penrose limit of some 
singular ``parent'' spacetime, wherein the singularity can be
timelike, spacelike, or null.  However, in taking the Penrose limit, any
null geodesic which terminates on these parent singularities, 
universally turns them into null singularities.  This constitutes the
second useful feature: the singularity in plane waves must be null, 
since it appears at a given $u$ value and extends in the $v$ direction.
For a discussion on singularities in plane waves see \host\ and 
more recently \mpz, \bgs.  
This universal behaviour of singularities in the Penrose limit 
is tantalizing and might hold some clues to resolving spacelike 
singularities. 
In string theory, one usually finds that timelike singularities are
generically much easier to ``resolve'' than spacelike ones (see however,
\shamit\ for an attempt at resolving spacelike singularities).
Typically, this is because it is easier to construct specific static
models which in the low energy SUGRA limit look singular, than to 
consider a fully dynamical set-up.
Recently, there has also been renewed interest in resolving 
null singularities, \cf, \refs{
\vijay, \lmsa, \albion, \horpol, \lmsb, \michal, \craps,\cornalbac}.

While the ``singularity resolution'' mechanism is model-specific,
the hope would be to take advantage of the universal features.
To be precise consider a scenario wherein two hitherto 
unrelated parent spacetimes, one with a spacelike $\M_s$ and other 
with a timelike singularity $\M_t$, lead to the same null singular 
plane wave background $\P$ upon taking appropriate Penrose limits. 
Assume furthermore that there is enough dynamical information 
at our disposal to resolve the timelike singularity in $\M_t$ 
to a smooth spacetime $\M_t^{{\rm res}}$.
In such an event one can ``resolve'' the singular plane wave 
spacetime $\P$ to $\P^{{\rm res}}$, by demanding that 
$\P^{{\rm res}}$ be obtained from $\M_2^{{\rm res}}$ by 
a Penrose limit analogous to the one used to get $\P$ from $\M_t$.
Running the same logic in the direction of $\M_s \to \P$, it 
might be possible to define a resolution of the spacelike 
singularity to a smooth spacetime $\M_s^{{\rm res}}$. 

Whereas in a previous work \nobh\ we asked the physical question 
of whether there can be black holes in pp-waves, 
here we have been concerned with finding the full causal structure of
pp-wave spacetimes.  
While our findings confirm our previous results,  
from the string theory point of view 
this may appear as a somewhat more esoteric question.
Causal structure is a classical notion, and our ideal point construction
is cast entirely within the classical theory of general relativity.
Within this setting, it is a 
fundamental concept, as it tells us which points in the spacetime
manifold are in causal contact with which other points, and therefore
the causal communication possible in the spacetime.
When the curvatures are small, and especially when the curvature 
invariants vanish as in the present case, we generally expect this
classical description to mimic what the ``fundamental'' objects in
string theory, such as strings, see.

However, the issue is more subtle, as hinted to by \eg\ the black hole
information paradox.  
What fundamentally nonlocal objects see is not quite the same as what
classical point particles see.
For instance, as pointed out by \prec,\gili\ in the context of AdS/CFT duality,
an event horizon need not pose a fundamental obstacle to holographically
extracting information from inside the black hole.\foot{
For various related discussions, see \eg\
\refs{\lmr,\kali,\maldacena,\jacobson}.}
These examples suggest that objects in string theory may see more
of the spacetime than allowed by classical causality.
On the other hand, one would expect that communication which is 
allowed by causality is also allowed within string theory; in this 
sense, the classical causal structure of a spacetime would provide
a ``lower bound'' on what can influence or be influenced by what.

At the same time, there are some indications that the causal structure as
seen by strings might be influenced by other fields in the supergravity 
background, such as the $NS-NS$ two-form field. Consider the case of the 
Nappi-Witten model \napwit\ as in \nwit. The metric in this solution is 
supported by a NS-NS field strength 
$H_{3}  = \l \, du \wedge dz^1 \wedge dz^2$.
When we study the world-sheet sigma model in light-cone gauge one can 
make a non-local transformation on the world-sheet to cast the massive
fields $\vz = (z^1,z^2)$ into free fields without a mass term. From this 
viewpoint, it appears as though the `effective' spacetime in terms of the 
world-sheet sigma model is flat space. However, as we have explicitly shown, 
the causal structure of \nwit\ is quite different from the corresponding 
10-dimensional Minkowski space and it in fact has a one-dimensional 
causal boundary. 

Perhaps a more important (but murkier) 
use of determining the causal structure 
of a spacetime has to do with possible holographic duals
of string theory on that background.
For example, 
the fact that the boundary of AdS is timelike allows for a natural
formulation of a dual Lorentzian description; in particular, the
global time in AdS spacetime is identical to the gauge theory notion of 
time. On a less rigorous footing is the example of the proposed dS/CFT
correspondence, \cf \andy,
where the fact that $\scri$ is spacelike suggests an Euclidean dual.
In the present case of pp-waves where we find a null one-dimensional 
causal boundary, we may ask what does this imply.
It has been speculated \kirit, \bena\
that the one-dimensionality of the boundary suggests a 
dual description in terms of pure quantum mechanics.

One has to be cautious, however, in nourishing such expectations, 
for the following reasons.
Firstly, although in AdS/CFT, the dual gauge theory is said to 
``live on the boundary'' of AdS, there is no reason for this notion
to be applicable elsewhere---even in cases where a holographic dual
does exist. In particular, in the spirit of Bousso's holographic 
bounds, \bousso,
the holographic screens need not correspond to the spacetime 
boundary.  (The fact that they do for AdS may be regarded as some 
support for this picture.)
From this standpoint, the causal structure of a spacetime
may well be entirely irrelevant.

Secondly, as we mentioned at the end of Section 3, the causal
boundary may not be the same thing as the conformal boundary.
In particular, the added ``endpoints'' of endless spacelike curves
need not correspond to the ideal points added to future/past-endless
causal curves.  In fact, even the topology may differ \mrtwo.
Unfortunately, 
from the lessons that one gleans from the AdS/CFT correspondence,
one might expect that the conformal boundary may be more important
than the causal one.

It would be interesting to understand these issues better, and 
in particular, to find whether there is (and if so, what is) the
stringy analog of the causal structure (\cf, \martineca, \martinecb, 
for some thoughts along these lines).
Unfortunately, it's not even clear what we would mean by such
an analog in the full string/M-theory:
So far, we have been used to dealing with string theory formulated
on some fixed background spacetime; however, one has to allow for the 
possibility that at the microscopic level spacetime itself
is a derived concept, at best to be treated on equal footing with
other collective field excitations. It would be very interesting to 
resolve these issues to enable us delve deeper into the 
workings of a quantum theory of gravity. 


\vskip 1cm

\centerline{\bf Acknowledgements}
It is a great pleasure to thank Chris Hillman, Petr Ho\v{r}ava, 
Gary Horowitz, and especially Don Marolf 
for illuminating discussions. 
We also would like to thank Don Marolf and Simon Ross for their 
comments on the manuscript.
We gratefully acknowledge the hospitality of Aspen Center for Physics,
where this project was initiated.
VH was supported by NSF Grant PHY-9870115, while MR acknowledges support 
from the Berkeley Center for Theoretical Physics and also partial support 
from the DOE grant DE-AC03-76SF00098 and the NSF grant PHY-0098840.

\appendix{A}{Geodesic equations and properties of pp-waves}

In this Appendix, we will summarize some useful facts about
plane waves \plane\ and pp-waves \pp, referred to at various stages in 
the paper.
Namely, we first derive the geodesic equations in plane waves,
and then list curvature tensors and geodesic equations
for the more general pp-waves.  The corresponding plane wave 
quantities can obviously be obtained by direct substitution.

\subsec{Geodesics in general plane waves}

Let us consider the general plane wave metric \plane.
Explicitly writing out the summations on transverse indices 
$i,j=1,\dots,d-2$ in $d$-dimensional spacetime 
(as we will do henceforth in this section to avoid confusion 
with components), the general plane wave metric is 
\eqn\splane{
ds^2 = -2 \, du \, dv - 
\sum_{i,j} f_{ij}(u) \, x^i x^j \, du^2 + 
\sum_i dx^i \, dx^i}

First consider the null geodesics in this spacetime.
Denote by 
\eqn\pa{
p^a = \ud \dua + \vd \dva + \sum_i \xid \dxa}
the tangent vector to the null geodesic.
Since $\dva$ is a Killing field, 
$p_a \, \dva = \ud$ is a constant of motion,
which we can set to $\ud \equiv 1$, so that $u$ acts as the affine 
parameter along the geodesic.
The null condition then implies
\eqn\nullp{
p_a \, p^a 
= - 2 \, \vd - \sum_{i,j} f_{ij} \, x^i \, x^j + \sum_i (\xid)^2 = 0}
where $f_{ij}$, $x^i$, and $v$ are implicitly functions of $u$, and 
$\dot{} \equiv {d \over du}$.
The Christoffel symbols for \splane\ can be easily found to be
\eqn\chr{
\Gamma^v_{\ uu} = {1 \over 2}  \sum_{i,j} \fd_{ij} \, x^i \, x^j, \ \ \ \ 
\Gamma^i_{\ uu} = \sum_j f_{ij} \, x^j = \Gamma^v_{\ ui}}
(with all other components vanishing),
so that the geodesic equations are
\eqn\geodv{
\vdd + {1 \over 2} \sum_{i,j} \fd_{ij} \, x^i \, x^j
     + 2 \, \sum_{i,j}  f_{ij} \, x^j \, \xid = 0}
\eqn\geodx{
\xdd^i + \sum_j f_{ij} \, x^j  = 0}
However, we can use the  first order constraint equation \nullp\
instead of \geodv, as can be checked by integrating \nullp\ and 
using \geodx\ to obtain \geodv.  Using \nullp\ and \geodx\ leads
to further simplification, since $\vd$ can be rewritten as a total 
derivative, so that we can solve for $v$ in terms of $x^i$:
\eqn\veq{
v = {1 \over 2} \, \sum_i x^i \, \xid + v_0}
where $v_0$ is an arbitrary integration constant which is fixed by the 
initial conditions.

\subsec{Overview of pp-waves}

For the pp-wave metric \pp, with $F(u,\vx) \equiv F(u, x^i)$,
\eqn\ppa{
 ds^2 = -2 \, du \, dv - F(u, \vx) \, du^2 +  \sum_i dx^i \, dx^i}
the Christoffel symbols are
\eqn\chrpp{
\Gamma^v_{\ uu} = {1 \over 2} \p_u F(u, \vx), \ \ \ \ \
\Gamma^i_{\ uu} = {1 \over 2} \p_i F(u, \vx)
 = \Gamma^v_{\ ui}}
(with all other components vanishing), 
where $\p_i \equiv {\p \over \p x^i}$, etc..
The Riemann tensor is given by
\eqn\riempp{
R_{uiuj} = {1 \over 2} \p_i \p_j F(u, \vx)}
with all other components vanishing, 
so that the Ricci tensor has the only nontrivial component
\eqn\Ruupp{
R_{uu} = {1 \over 2} \nabla_x^2 F(u, \vx)}
where $ \nabla_x^2$ is the transverse Laplacian.
The Ricci scalar then vanishes, $R=0$, so the Einstein tensor 
is identical to the Ricci tensor.
For completeness, we note that the Weyl tensor is given by
\eqn\weylpp{
C_{uiuj} = {1 \over 2} \( \p_i \p_j F(u, \vx) 
            - {1 \over d-2} \, \delta_{ij} \,
              \sum_k \p_k^2 F(u, \vx) \) }

Now, let us consider null geodesics in the pp-wave background.
In the above coordinates, these are given by
\eqn\geodxpp{
\xdd^i + {1 \over 2} \p_i F(u, \vx)  = 0}
\eqn\geodvpp{
\vd = - {1 \over 2} F(u, \vx) + {1 \over 2} \sum_i( \xid)^2 }
Note that unless $F(u, \vx)$ takes the plane wave form $f_{ij}(u) \, x^i x^j$,
$\vd$ is not a total derivative, so we can't write a general
formula for $v$ analogous to \veq, without first solving the 
$x$-equations \geodxpp.

\appendix{B}{Null geodesics in vacuum pp-waves}

We present numerical solutions to geodesic equations in 
vacuum pp-wave backgrounds to illustrate our point that these spacetimes 
are geodesically incomplete. We will deal with two main 
examples in the following. 

\vskip3mm
\noindent
{\bf Ex.1.} The first example is the spacetime
\eqn\fdvpprfour{
ds^2 = -2 \, du \, dv - \[ 35 \, z^4 - 30 z^2 \, (z^2 + x^2 + y^2) + 
3 \, (z^2 + x^2 + y^2)^2 \] \, du^2  + dx^2 + dy^2 + dz^2
}
This spacetime is a vacuum pp-wave in five spacetime dimensions. It 
can be cast into the familiar form of \pparad, by writing the 
function $F(z,x,y)$ appearing in $g_{uu}$ as $r^4 P_4(\cos \theta)$, 
where $P_4(x)$ is the Legendre polynomial of degree $4$.

The geodesic equations \geodxpp\ in the background \fdvpprfour\ are
\eqn\geovfive{\eqalign{
& \ddot{x} + \( 6  \, x(u)^3 \, + 6  \, x(u)  \, y(u)^2 
- 24  \, x(u)  \, z(u)^2 \) = 0 \cr
& \ddot{y} + \( 6  \, y(u)^3 \, + 6  \, x(u)  \, y(u)^2 - 24 \,
y(u) \, z(u)^2 \) = 0 \cr
& \ddot{z} + \( 16  \, z(u)^3 - 24 \, z(u) \( x(u)^2 + y(u)^2 \) \) = 0 
}}

\ifig\Figvaca{Geodesics $\{z(u),x(u),y(u)\}$ in the background \fdvpprfour.}
{\epsfxsize=8cm \epsfysize=6cm \epsfbox{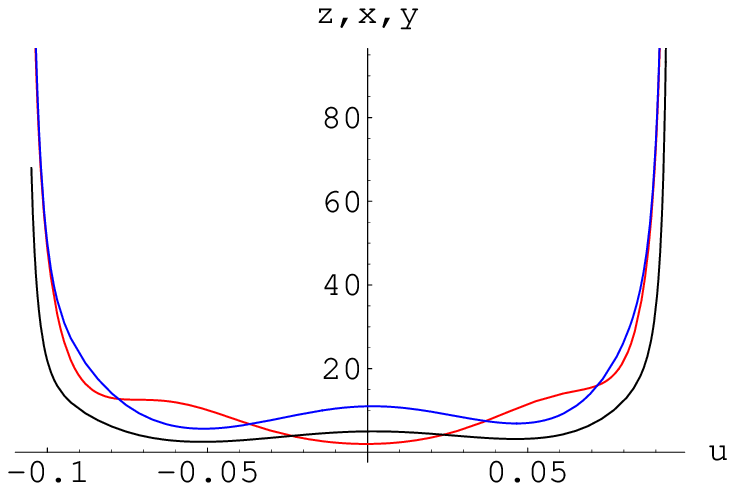}}

We have numerically integrated the equations \geovfive\ and found that 
the geodesics typically reach infinite values of the coordinates in 
finite affine parameter. One such geodesic is illustrated in Fig.7.

\vskip3mm
\noindent
{\bf Ex.2.} Our next example is the pp-wave background 
with $F(u,r\to \infty,\Om) \to 0$, discussed in 
{\bf PP3}. Specifically, the spacetime background we 
consider is 
\eqn\fdppzeroappb{
ds^2 = -2 \, du \, dv - {1 \over r^4} \, \(5 \cos^3 \theta - 3 \cos \theta \) 
 \, du^2 + dr^2 + r^2\, \(d\theta^2 + \sin^2 \theta d \phi^2 \) }
We will first take a short detour to 
determine the geodesics in a general 5-dimensional 
spacetime, and then proceed to solve them for the particular
spacetime \fdppzeroappb\ at hand.

Since to determine the causal structure, we would like to 
consider radial curves, let us consider the null geodesics 
in the spherical coordinates.
To that end, we start with the metric \pparad.
If $F$ has no angular dependence, $F(u,r,\Om) = F(u,r)$, then
we obviously have
$\ddot{r} + {1 \over 2} \p_r F(u, r)  = 0$ and
$2 \vd = - F(u, r) + \dot{r}^2$, with $\dot{\Om} = 0$.
However, if $F$ has some angular dependence, these no longer
describe null geodesics; in other words, the geodesics cannot remain
radial.

To illustrate this, let us consider a 5-dimensional spacetime, $d=5$.
For $F = F(u,r,\theta,\phi)$, the geodesic equations are
$$\ddot{v} + {1 \over 2} \dot{F} +  \p_r F \, \dot{r}
+  \p_{\theta} F \, \dot{\theta} +  \p_{\phi} F \, \dot{\phi} = 0$$
$$\ddot{r} + {1 \over 2} \p_r F -r 
\[ \dot{\theta}^2 + \sin^2\theta \, \dot{\phi}^2 \] = 0$$
$$\ddot{\theta} + {1 \over 2r^2} \p_{\theta} F
+ {2 \over r} \, \dot{r} \, \dot{\theta} 
- \sin\theta \cos\theta \, \dot{\phi}^2 = 0$$
\eqn\geodpp{
\ddot{\phi} + {1 \over 2r^2 \, \sin^2\theta} \p_{\phi} F
+ {2 \over r} \, \dot{r} \, \dot{\phi} 
+ 2 {\cos\theta \over \sin\theta} \, \dot{\theta} \, \dot{\phi} = 0}
As before, we can exchange the $v$-equation for 
 the first order null constraint,
\eqn\nullpp{
2\vd= -F+ \dot{r}^2 + r^2 
\( \dot{\theta}^2 + \sin^2\theta \, \dot{\phi}^2 \)}
as can be easily checked by integrating \nullpp\ and substituting
in \geodpp.

\ifig\Figvacb{Numerical solutions to geodesics $\{r(u),\theta(u) \}$ 
for the spacetime background \fdppzero. The plot on the left shows 
geodesics which exist for the full range of the affine parameter $u$. 
The plot on the right shows geodesics which diverge at finite 
values of the affine parameter.}
{\epsfxsize=12cm \epsfysize=5cm \epsfbox{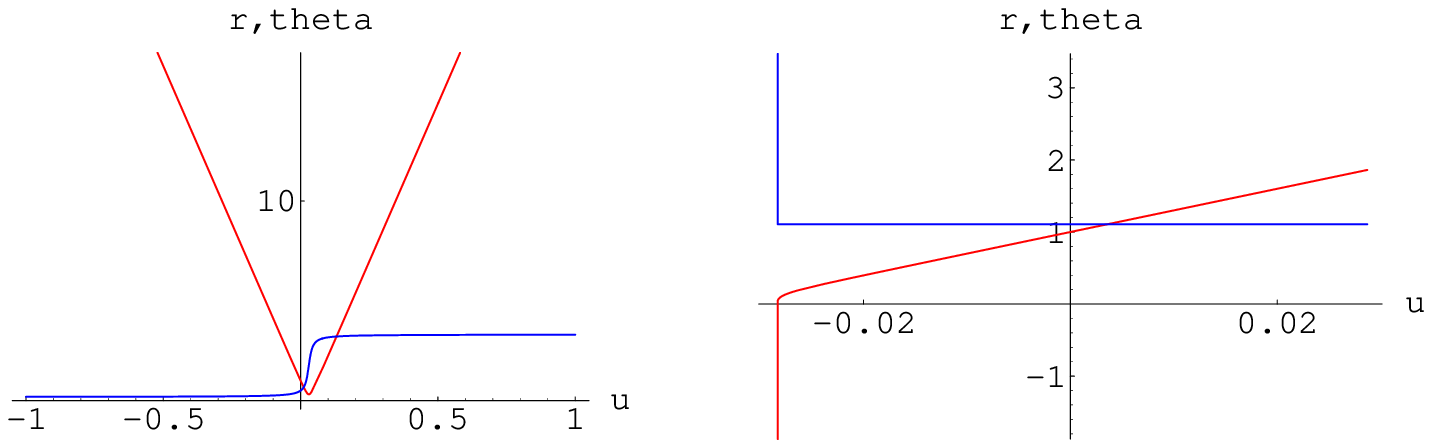}}

Specializing to the background \fdppzero\ where $F(r, \theta) = 
{1 \over r^4} \, \(5 \cos^3 \theta - 3 \cos \theta \) $, we find that the 
geodesic equations reduce to
\eqn\fdppzerogeod{\eqalign{
& \ddot{r } - {2 \over r(u)^5 } \,  \(5 \cos^3 \theta(u) - 3 \cos \theta(u) \)
- r(u) \, \dot{\theta}(u)^2 = 0 \cr
& \ddot{\theta} - {1 \over 2 r(u)^6} \(15 \cos^2 \theta(u) - 3\) 
\sin \theta(u) + {2 \over r(u)} \, \dot{r} \, \dot{\theta} = 0
}}

The results of numerically solving \fdppzerogeod\ are plotted in Fig.8. 
As mentioned in {\bf PP3}, there are two types of behaviour. The 
first plot on the left, demonstrates that there are indeed geodesics which 
are well behaved for all values of the affine parameter. However, from the 
point of view of determining the causal structure, they don't help us 
reach large negative values of $v$. This happens because $\dot{v}$ is 
determined in terms of $\dot{r}^2$ and $\dot{\theta}^2$, since the 
function $F(u,r,\theta) \sim {1 \over r^4} \to 0$ for sufficiently large $r$.
This rapid approach to flat space causes the geodesics to tend towards 
$v \to + \infty$ rather than $ v \to -\infty$.  In the second plot, on the 
right of Fig.8, we show that this spacetime is geodesically incomplete. The 
geodesics in this case diverge at finite values of the affine parameter.

%
%

\listrefs
\end